\documentclass[onecolumn,amsmath,amssymb,nofootinbib,12pt]{article}
\usepackage{jheppub}
\usepackage{ifpdf}
\usepackage{physics}
\usepackage{graphicx,subcaption}
\usepackage{amsfonts}
\usepackage{amsmath}
\usepackage{amssymb}
\usepackage{epsfig}
\usepackage{pdfpages}
\usepackage{graphicx,epstopdf}
\usepackage[makeroom]{cancel}
\usepackage{hyperref}
\hypersetup{pdftex,colorlinks=true,allcolors=blue}
\usepackage{array}
\usepackage{ulem}
\newcommand{\RN}[1]{%
  \textup{\uppercase\expandafter{\romannumeral#1}}%
}

\newcommand{\maxx}{\mt{max}}

\newcommand{\RNo}{\mt{RNA}}

\newcommand{\be}{\begin{equation}}
\newcommand{\ee}{\end{equation}}
\newcommand{\bea}{\begin{eqnarray}}
\newcommand{\eea}{\end{eqnarray}}
\newcommand{\beq}{\begin{equation}}
\newcommand{\eeq}{\end{equation}}
\newcommand{\beqa}{\begin{eqnarray}}
\newcommand{\eeqa}{\end{eqnarray}}
\newcommand{\beqar}{\begin{eqnarray*}}
\newcommand{\eeqar}{\end{eqnarray*}}

\def\bal#1\eal{\begin{align}#1\end{align}}
\def\r{\rightarrow}

\newcommand{\labell}[1]{\label{#1}} 

\newcommand{\eg}{{\it e.g.,}\ }
\newcommand{\ie}{{\it i.e.,}\ }
\newcommand{\reef}[1]{(\ref{#1})}
\newcommand{\ssc}{\scriptscriptstyle}
\newcommand{\mt}[1]{\textrm{\tiny #1}}

\def\f {\frac}

\def\qq {\quad\quad}

\def\no{\nonumber\\}

\renewcommand{\l}{\ell}

\def\S{\Sigma}
\newcommand{\cv}{{\cal C}_\mt{V}}
\newcommand{\ca}{{\cal C}_\mt{A}}

\newcommand{\ctL}{\ell_\mt{ct}}

\newcommand{\qt}{q_{\ssc T}}
\newcommand{\qtx}{q_{\ssc T,ext}}
\newcommand{\ap}{\gamma}

\newcommand{\al}{\alpha}
\newcommand{\gaa}{\gamma_\alpha}
\newcommand{\lct}{\ell_\mt{ct}}

\usepackage{xspace}

\begin{document}

\hfill \begin{minipage}{3.5cm} {UT-Komaba/18-6}\end{minipage}
\preprint{arXiv:1901.nnnnn [hep-th]}
\title{Holographic Complexity Equals Which Action?}

\author[a]{Kanato Goto,}
\author[b, c]{Hugo Marrochio,}
\author[b]{Robert C Myers,}
\author[b,c]{Leonel Queimada}
\author[b]{and Beni Yoshida}
\affiliation[a]{The University of Tokyo, 
 Komaba, Meguro-ku, Tokyo 153-8902 Japan}
\affiliation[b]{Perimeter Institute for Theoretical Physics, Waterloo, ON N2L 2Y5, Canada}
\affiliation[c]{Department of Physics $\&$ Astronomy,
University of Waterloo,\\ Waterloo, ON N2L 3G1, Canada}

\emailAdd{kgoto@hep1.c.u-tokyo.ac.jp,
hmarrochio@perimeterinstitute.ca,
rmyers@perimeterinstitute.ca,
lquintaqueimada@perimeterinstitute.ca,
byoshida@perimeterinstitute.ca}

\date{\today}

\abstract{We revisit the complexity$=$action proposal for charged black holes. We investigate the complexity for a dyonic black hole, and we find the surprising feature that the late-time growth is sensitive to the ratio between electric and magnetic charges. In particular,  the late-time growth rate vanishes when the black hole carries only a magnetic charge. If the dyonic black hole is perturbed by a light shock wave, a similar feature appears for the switchback effect, \eg it is absent for purely magnetic black holes. We then show how the inclusion of a surface term to the action can put the electric and magnetic charges on an equal footing, or more generally change the value of the late-time growth rate. Next, we investigate how the causal structure influences the late-time growth with and without the surface term for  charged black holes in a family of Einstein-Maxwell-Dilaton theories. Finally, we connect the previous discussion to the complexity=action proposal for the two-dimensional Jackiw-Teitelboim theory. Since the two-dimensional theory is obtained by a dimensional reduction from Einstein-Maxwell theory in higher dimensions in a near-extremal and near-horizon limit, the choices of parent action and parent background solution determine the behaviour of holographic complexity in two dimensions. }

\maketitle


\section{Introduction}\label{sec:Intro}
Recent years have witnessed that quantum information theoretic notions shed new light on deep conceptual puzzles in the AdS/CFT correspondence, and also provide useful tools to study the dynamics of strongly-coupled quantum field theories, \eg \cite{Ryu:2006bv,Ryu:2006ef, EntBook, Hubeny:2007xt, HoloEntEntropy,Swingle:2009bg, Myers:2010tj, Blanco:2013joa,Dong:2013qoa, Faulkner:2013ica, Almheiri:2014lwa,Pastawski:2015qua}. One striking, yet mysterious, entry to the gravity/information dictionary is the concept of quantum circuit complexity: the size of the optimal circuit which prepares a target state from a given reference state with a set of ``simple'' gates  \cite{Aaronson:2016vto,johnw,Susskind:2018vql,Susskind:2018pmk}.  The concept of holographic complexity naturally emerges from considerations on the bulk causality in the AdS/CFT correspondence \cite{Susskind:2014rva}. For instance, holographic complexity is expected to be a useful diagnostic for late-time dynamics and in particular, the interior of a black hole since it continues to increase even after the boundary theory has reached thermal equilibrium. In addition, complexity is sensitive to perturbations of the system, namely the physics of scrambling, \ie even tiny perturbations to the system have a measurable effect on the complexity \cite{Susskind:2014rva}.

The original proposal for holographic complexity, known as the complexity$=$volume (CV) conjecture, asserts that  \cite{Susskind:2014rva, Stanford:2014jda}
\begin{equation}
\cv(\S) =\ \mathrel{\mathop {\rm
max}_{\scriptscriptstyle{\S=\partial \mathcal{B}}} {}\!\!}\left[\frac{\mathcal{V(B)}}{G_N \, L}\right] \, ,\labell{defineCV}
\end{equation}
where the boundary state lives on the time slice $\Sigma$ and then $\mathcal V (\mathcal{B})$ is the volume of codimension-one bulk surfaces $\mathcal{B}$ anchored to this boundary time slice. To produce a dimensionless quantity, the volume is divided by  Newton's constant $G_N$ and the AdS length $L$.\footnote{A more sophisticated approach to choosing the latter scale was described in \cite{ted}.} A second proposal, known as the complexity$=$action (CA) conjecture, relates the complexity of the boundary state to the gravitational action evaluated for a particular region of the bulk spacetime  \cite{Brown1, Brown2}
\beq
\ca(\S) =  \frac{I_\mt{WDW}}{\pi\, \hbar}\,. \labell{defineCA}
\eeq
Here, the subscript WDW indicates that the action is calculated on the so-called Wheeler-DeWitt patch, which corresponds to the causal development of any of the above bulk surfaces $\mathcal B$ anchored on $\Sigma$.

The study of holographic complexity is actively developing in two related directions. The first is to explore the properties of the new gravitational observables which play a role in the CV and CA conjectures and the implications of these conjectures for complexity in the boundary theory, \eg see \cite{Susskind:2014rva, Stanford:2014jda, ted, Brown1, Brown2, Susskind:2014moa, Susskind:2015toa, Roberts:2014isa, Cai:2016xho, NullBound, Reynolds:2016rvl, Format, Cai:2017sjv, diverg, CompLove, 2LawComp, Bridges, Moosa, Growth, fish, brian, Alishahiha:2018tep, An:2018xhv, Mahapatra:2018gig, Vad1, Vad2, Zhao:2017isy, Flory:2017ftd, Abt:2017pmf, Abt:2018ywl,Fu:2018kcp, Agon:2018zso, Barbon:2018mxk, Cooper:2018cmb, Numasawa:2018grg,Susskind:2018fmx, Susskind:2018tei, Brown:2018kvn, fish2}.  In particular, we note that a number of new proposals have been made for the holographic dual of complexity in the boundary theory. For example, one new proposal is known as the complexity$=$(spacetime volume) (or CV2.0) conjecture, which suggests that the bulk dual of complexity is the spacetime volume of the WDW patch \cite{fish2}. Further, more recently, a relation was conjectured between momentum of an infalling object in the bulk radial direction and complexity of the corresponding time-evolved operator on the boundary \cite{Susskind:2018tei, Brown:2018kvn}.
A second direction of investigation has been to understand the concept of circuit complexity for quantum field theory states, in particular for states in a strongly coupled CFT, \eg see \cite{EuclideanComplexity1, EuclideanComplexity2, EuclideanComplexity3, EuclideanComplexity4, Reynolds:2017lwq, Roberts:2016hpo, koji, qft1, qft2, qft3, Coherent, TFD, Bhattacharyya:2018bbv, Caputa:2018kdj, Belin:2018fxe, Belin:2018bpg, Balasubramanian:2018hsu, Takayanagi:2018pml} 
Developing a proper understanding of complexity in the boundary theory is essential to properly test the various holographic proposals and ultimately, to produce a derivation of one (or more) of these conjectures.

Our motivation for the present paper was to understand holographic complexity (and in particular, the CA proposal) in the two-dimensional Jackiw-Teitelboim (JT) model of dilaton-gravity \cite{Jackiw:1984je, Teitelboim:1983ux}. Recently, there has been a great deal of interest in the JT model, as it emerges in the holographic description of the Sachdev-Ye-Kitaev (SYK) model in a particular low energy limit, where the system acquires an emergent reparametrization invariance \cite{AP, PhysRevLett.70.3339, Kitaev, Kitaev2, Plochinski:2016, Maldacena:2016hyu, Maldacena:2016upp, EMV, Jensen:2016pah, Almheiri:2016fws, Sarosi:2017ykf, Nayak:2018qej}. Furthermore, the JT gravity exhibits the late-time behavior of the spectral form factor which are natural from the perspective of random matrix theory, \eg \cite{Garcia:2016,Cotler:2016}. As such, the JT model should be an ideal platform to study the complexity in various dynamical settings and investigate further the implications of holographic complexity. However, our initial calculation of holographic complexity in the JT model using the CA proposal \reef{defineCA} produced the surprising result that the growth rate vanishes at late times --- see section \ref{sec:JT}. This result, of course, in tension with our common expectations for complexity. It can be argued quite generally that at late times, the  complexity should continue to grow with a rate given by  \cite{Susskind:2014rva,Susskind:2014moa}
\begin{equation}
\frac{d{\mathcal C}}{dt} \sim S\,T \,, \label{general}
\end{equation}
where the entropy $S$ gives an account of the number of degrees of freedom while the temperature $T$ sets the scale for the rate at which new gates are introduced. Further since the JT model is supposed to capture the physics of the SYK model, which exhibits maximal chaos, we would certainly expect the complexity should increase as fast as it possibly can. Rather than considering the CA prescription of holographic complexity, one can also examine the CV proposal \reef{defineCV} in this setting and in this case, we found the extremal volume (\ie the length of the geodesic connecting the boundary points defining $\Sigma$) continues to grow at a constant rate for arbitrarily late times. 

This apparent failure of the CA proposal motivated us to re-examine holographic complexity for charged black holes in higher dimensions 
since the JT model can be derived from an appropriate dimensional reduction, \eg \cite{Sen:2007qy,Kunduri:2007vf, Kunduri:2013ana, Castro:2008ms, Nayak:2018qej, Almheiri:2016fws, Sarosi:2017ykf}. In particular, JT dilaton-gravity describes the near-horizon physics of certain near-extremal black holes in higher dimensions. Previous studies of holographic complexity of charged black holes, \eg \cite{Brown1, Brown2,Cai:2017sjv,Growth}, had not shown any odd behaviour for the CA proposal. However, with hindsight, we note that all of these investigations involved electrically charged black holes, whereas the usual dimensional reduction made to derive the JT model involves black holes carrying a magnetic charge, \eg \cite{Sen:2007qy, Nayak:2018qej}. Our first calculation in the following is to examine holographic complexity for a dyonic black hole (in four dimensions) with both electric and magnetic charges. In this case, even if the geometry is held fixed, we find that the complexity growth rate is very sensitive to the ratio between the two types of charge. In particular, if the black hole is purely magnetic, we find that CA proposal yields a vanishing growth rate at late times, and further, that the switchback effect is absent. Of course, this vanishing matches our result for the JT model, which would arise in the dimensional reduction of these magnetic black holes.

However, there is a boundary term involving the Maxwell field, which one might add to the gravitational action. This term arises naturally in the context of black hole thermodynamics \cite{char0} (see also \cite{char1,char2}) when defining different thermodynamics ensemble, \ie a canonical ensemble with fixed charge, as opposed to a grand canonical ensemble with fixed chemical potential. We find that with the CA proposal, the holographic complexity is also sensitive to the introduction of this Maxwell surface term. In particular, the late-time growth rate is nonvanishing for magnetic black holes with this surface term, while tuning the coefficient of the surface term can yield a vanishing growth rate in the electrically charged case.  Given these results, we are then lead to re-examine the dimensional reduction in the presence of the Maxwell surface term and the behaviour of the corresponding holographic complexity for the JT model and for a related ``JT-like" model, derived from the reduction of  electrically charged black holes.

To better understand  the vanishing of the complexity growth for the magnetic black holes, we might also ask whether this result is special to the Einstein-Maxwell theory. Alternatively, the question can be phrased as which features of the corresponding Reissner-Nordstrom-AdS black holes are important in controlling the behaviour of the holographic complexity for the CA proposal. As a step in this direction, we also investigate holographic complexity for  charged black holes in a family of four-dimensional Einstein-Maxwell-Dilaton theories. Holographic complexity of Einstein-Maxwell-Dilaton theories has been previously studied for several models \cite{Cai:2017sjv, brian, Alishahiha:2018tep, An:2018xhv, Mahapatra:2018gig}. In these theories, the Maxwell field is coupled to a scalar field (the dilaton) and as a result, the charged black holes also carry ``scalar hair.'' With the dilaton excited in these solutions,  the nature of the spacetime singularities and the casual structure of the corresponding black holes can be modified. Hence we can investigate to what extent these changes to the spacetime geometry modify the behaviour of the holographic complexity. Our conclusion will be that the causal structure of the spacetime geometry is the essential feature leading to the unusual (\ie vanishing) late-time growth rate with the CA proposal.

The remainder of our paper is organized as follows: In section \ref{sec:RNBHS}, we study the CA proposal for dyonic black holes carrying both electric and magnetic charges in four bulk dimensions. We first show how for a fixed geometry, the complexity rate of change is sensitive to the ratio between electric and magnetic charges. We also show how the inclusion of the Maxwell surface term to the action can also have a dramatic effect on the late-time growth rate for the CA proposal. In addition, we also briefly investigate the switchback effect by injecting small shockwaves into the dyonic black hole. In section \ref{sec:Dila}, we investigate holographic complexity for charged black holes in a family of Einstein-Maxwell-Dilaton theories.  In section \ref{sec:JT}, we return to holographic complexity for two-dimensional black holes. In particular, we show that the late-time growth rate vanishes for the JT model, but that this situation can be ameliorated if the Maxwell surface term is included in reduction from four to two dimensions. We summarize our findings and consider their implications in section \ref{sec:Discussion}, as well as discussing some possible future directions.  We leave certain technical details to the appendices. In appendix \ref{App:Shock}, we describe in more detail the calculations of the holographic complexity in the dyonic shock wave geometries. In appendix \ref{MBAapp}, we comment on subtleties concerning the evaluation of the Maxwell surface term when magnetic charges are present.

As this project was nearing its completion, we became aware of  \cite{Brown:2018bms}, which has significant overlap with the present paper. We also add that an independent approach to understanding holographic complexity for the JT model recently appeared in \cite{Alishahiha:2018swh, Akhavan:2018wla}.

\section{Reissner-Nordstrom Black Hole}\label{sec:RNBHS}
In this section, we investigate applying the complexity=action (CA) conjecture \cite{Brown1,Brown2} to evaluate the holographic complexity of the dyonic Reissner-Nordstrom black hole, while focusing on the Einstein-Maxwell theory in four bulk  dimensions, \ie $d=3$ for the boundary theory. 
These results are easily extended to general dimensions, if one also couples the gravity theory to a ($d$--2)-form potential field (\ie the Hodge dual of the one-form Maxwell potential). Our main objective is to understand the effect of a new boundary term associated with the Maxwell field. As mentioned in the introduction, we will find that although this boundary term does not modify the field equations, it has a strong influence on the action of the Wheeler-DeWitt (WDW) patch. Hence we must ask which choices (for the coefficient of this term) yield a WDW action which produces the behaviours expected of holographic complexity.

We divide the action for four-dimensional Einstein-Maxwell theory in terms of the usual Einstein-Hilbert and Maxwell actions, as well as various possible surface terms
\begin{equation}\label{TOTaction}
I_\mt{tot} = I_\mt{EH} + I_{\mt{Max}}+ I_\mt{surf} + I_{\mt{ct}} + I_{\mu \mt{Q}} \, ,
\end{equation}
where first two terms are integrated over the bulk of the manifold of interest  
\begin{equation}\label{bulk}
\begin{split}
I_\mt{EH} = &\ \frac{1}{16 \pi G_N} \int_\mathcal{M} d^{4} x\, \sqrt{-g} \left(\mathcal R + \frac{6}{L^2}\right)\,, \\
I_{\mt{Max}} =&\, - \frac{1}{4 g^2} \int_{\mathcal{M}} d^{4} x\, \sqrt{-g} \, F_{\mu \nu} F^{\mu \nu} \, .
\end{split}
\end{equation}
The next term $I_{surf}$ contains various surface terms needed to make the variational principle well-defined for the metric, 
\begin{equation}\label{surface}
\begin{split}
I_\mt{surf} = 
&\ \frac{1}{8\pi G_N} \int_{\mathcal{B}} d^3 x \sqrt{|h|} K + \frac{1}{8\pi G_N} \int_\Sigma d^{2}x \sqrt{\sigma} \eta
\\
&  + \frac{1}{8\pi G_N} \int_{\mathcal{B}'}
d\lambda\, d^{2} \theta \sqrt{\gamma} \kappa
+\frac{1}{8\pi G_N} \int_{\Sigma'} d^{2} x \sqrt{\sigma} a \, ,
\end{split}
\end{equation}
This contains the usual Gibbons-Hawking-York term \cite{York, GH} for timelike and spacelike boundary segments, the Hayward terms \cite{Hay1, Hay2} for intersections of these segments, and  the surface and joint terms introduced in \cite{NullBound} for null boundary segments --- see \cite{NullBound} for a complete discussion.  The null surface counterterm,  
 \begin{equation}\label{Counterterm}
I_\mt{ct} = \frac{1}{8 \pi G_N} \int_{\mathcal{B}'} d \lambda \, d^{2} \theta \sqrt{\gamma}\, \Theta \,\log \left(\ctL \Theta \right) \, ,
\end{equation}
is not needed for the variational principle, but it was introduced
in \cite{NullBound} to ensure reparametrization invariance on the null boundaries. Further, it was shown with a careful study of shock wave geometries in \cite{Vad1,Vad2} that this surface term must be included on the null boundaries of the WDW patch if the CA proposal is to reproduce the expected properties of complexity.

The final contribution in eq.~\reef{TOTaction}
is a boundary term for the Maxwell field
\begin{equation}\label{mba0}
I_{\mu\mt{Q}}= \frac{\ap}{g^2} \int_{\partial \mathcal{M}} d \Sigma_{\mu} \, F^{\mu \nu} \, A_{\nu} \, .
\end{equation}
While introducing this boundary term does not change the equations of motion, it does change the nature of the variational principle of the Maxwell field. That is, it changes the boundary conditions that must be imposed for consistency of the variational principle. We will also find that it modifies the WDW action, but we reserve a complete discussion of this term for section \ref{sec:MBA}. 

For the calculations which are immediately following, we will drop the Maxwell boundary term \reef{mba0} by setting the parameter $\ap=0$. That is, we examine the holographic complexity working with the action
\beq
I_\mt{0}=I_\mt{tot}(\ap=0)\,.
\label{Inot}
\eeq

\begin{figure}
\centering
\includegraphics[scale=0.35]{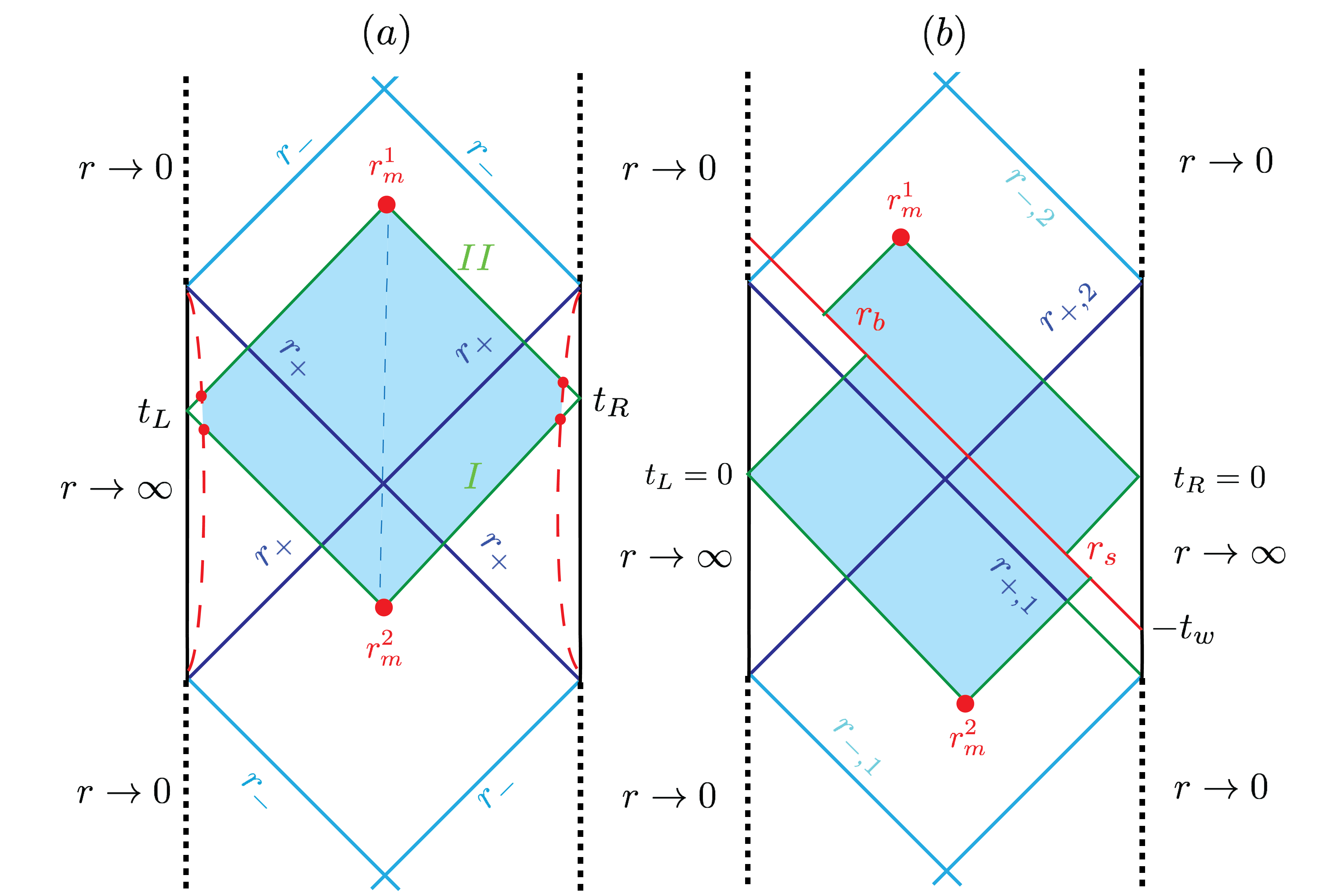}
\caption{(a) Penrose diagram for the Reissner-Nordstrom-AdS black hole \reef{RNBas}. The nonextremal black holes have an outer event horizon at $r=r_{+}$ and an inner Cauchy horizon $r=r_{-}$. The shaded blue region corresponds to a typical WDW patch anchored to symmetric boundary time slices with $t_L = t_R = t/2$. (b) Penrose-like diagram of an Reissner-Nordstrom-AdS black hole with a shock wave inserted on the right boundary at the boundary time $t_R=- t_w$ -- see eq.~\reef{shocker}. In order to not tilt the asymptotic boundary after the shock wave, we adopt the Dray-t'Hooft prescription that the null geodesics crossing the collapsing shock wave shifts. }
\label{PenroseHigherCharged}
\end{figure}

With this action, we apply the CA proposal to study the holographic complexity for a spherically symmetric {\it dyonic} Reissner-Nordstrom-AdS black hole (with $d=3$ boundary dimensions). The spacetime geometry is described by the following metric,
\begin{align}
& d s^2 = -f_{\RNo}(r) d t^2 + \frac{d r^2}{f_{\RNo}(r)} + r^2\, (d \theta^2 + \sin ^2 \theta d \phi^2 ) \nonumber \\
&\quad{\rm with} \qquad f_{\RNo}(r) = \frac{r^2}{L^2}+1 - \frac{\omega }{r} + \frac{q_e^2 + q_m^2}{r^2} \, , \label{RNBas}
\end{align}
where $L$ is the AdS length, and $\omega$ is a parameter proportional to the mass. A Penrose diagram showing the causal structure is shown in figure \ref{PenroseHigherCharged}(a), with an outer horizon $r_{+}$ and inner Cauchy horizon $r_{-}$ (defined by $f_{\RNo}(r_{\pm}) = 0$). The mass, entropy and temperature are then given by
\begin{equation}
M = \frac{ \omega }{2 G_N}\, , \qquad S= \frac{ \pi}{G_N} r_+^{2} \, ,\qquad T = \frac{1}{4 \pi} \frac{\partial f_{\RNo} }{\partial r} \bigg{|}_{r = r_+} \, . \label{MST}
\end{equation}
As indicated above, the black hole carries both electric and magnetic charges. The corresponding Maxwell field strength and vector potential can be written as
\begin{align}
&A= \frac{g}{\sqrt{4 \pi G_N}} \,  \left( q_m(1- \cos \theta) \, d \phi +\left( \frac{q_e}{r_{+} } - \frac{q_e}{r } \right) d t  \right)\nonumber \, , \\
&F = \frac{g}{\sqrt{4 \pi G_N}} \left( \frac{q_e}{r^2} \, d r \wedge dt  + q_m \sin \theta\, d \phi \wedge d \theta \right)\, . \label{FieldDyon}
\end{align}
where $q_e$ and $q_m$ denote the electric and magnetic charges. 

Following the conventions of \cite{Growth}, we write the tortoise coordinates for the black hole spacetime \reef{RNBas}, as 
\begin{equation}
r_{\RNo}^{*} (r) = - \int_{r}^{\infty} \frac{d \tilde r}{f_{\RNo} (\tilde r)} \, ,\label{rstar}
\end{equation}
such that $\lim_{r \rightarrow \infty} r_{\RNo}^{*} (r) = 0$. The Eddington-Finkelstein coordinates,  $v$ and $u$, for ingoing and outgoing rays (from the right boundary), respectively, are given by
\begin{equation}
v = t + r^{*} (r) \, , \qquad \qquad u= t - r^* (r) \, . \label{uv}
\end{equation}

\subsection{Complexity Growth} \label{subsec:CompDyon}

Next, we evaluate the growth rate of the holographic complexity for the dyonic black hole \reef{RNBas}. This analysis reveals the puzzling feature that despite the fact that magnetic and electric charges are interchangeable at the level of the equations of motion, the complexity growth in the CA proposal \reef{defineCA} seems to be sensitive to the nature of the charge. In the following, we provide salient points in the calculation and we refer the interested reader to \cite{Growth} for further details. 

Following \cite{Growth}, we anchor the WDW patch symmetrically on the left and right asymptotic boundaries with $t_L = t_R = t/2$. A typical WDW patch is illustrated in the Penrose diagram in figure \ref{PenroseHigherCharged}(a). The time evolution of the WDW patch can be encoded in the time dependence of points where the null boundaries intersect in the bulk, \ie the future boundaries meet at $r=r_m^1$ (and $t=0$) while the past boundaries, at  $r=r_m^2$ (and $t=0$), as shown in figure \ref{PenroseHigherCharged}(a). The position of these meeting points is determined by \cite{Growth}
\begin{equation}
\frac{t}{2} - r_{\RNo}^{*} (r_m^1) =0 \, , \qquad \qquad \frac{t}{2} + r_{\RNo}^{*} (r_m^2) =0  \, , \label{MeetPntRN}
\end{equation}
and then the rate at which these positions change is simply given by
\begin{equation}
\frac{d r_m^1}{d  t} = \frac{f_{\RNo} ( r_m^1)}{2} \, , \qquad \qquad \frac{d r_m^2}{d t} = - \frac{f_{\RNo} ( r_m^2)}{2}   \, . \label{MeetPntDerRN}
\end{equation}

\subsubsection*{Bulk contribution}

We start by evaluating the time derivative of the two bulk terms in eq.~\eqref{bulk}. With the Reissner-Nordstrom geometry \eqref{RNBas} and the Maxwell field \reef{FieldDyon}, these terms yield
\beq
I_{\text{bulk}} =I_\mt{EH}+I_\mt{Max}= \frac{1}{4 G_N} \int_{\mt{WDW}}\!\!\! dr\,dt \, r^2  \left( - \frac{6}{L^2} + \frac{2  \, ( q_e^2- q_m^2)}{r^4} \right) \, , 
\label{BulkIntegrand}
\eeq
where we have used the trace of Einstein equations: $\mathcal{R} = - \frac{12}{L^2}$. Notice that in the Maxwell contribution (\ie the second term in the integrand), the electric and magnetic charges appear with opposite signs! This fact is directly related to the vanishing of the late time rate of complexity for magnetic black holes, as we will see below.
Following \cite{Growth, CompLove}, the time derivative of the bulk action reduces to the difference of terms evaluated at the future and past meeting points,  
\begin{equation}
\frac{d I_{\text{bulk}} }{d t} =   \frac{1}{2 G_N} \left[ \frac{r^3}{L^2} + \frac{q_e^2- q_m^2}{r} \right]_{r_m^2}^{r_m^1} \, . \label{BulktDer}
\end{equation}

\subsubsection*{Joint contributions}

As shown in figure \ref{PenroseHigherCharged}(a), the WDW patch is cut off by a UV regulator surface at some large $r=r_\mt{max}$. However, the boundary contributions coming from this time-like surface segment and the corresponding joints yield a fixed constant, \ie they do not contribute to the time derivative of the action. Further, with affinely-parametrized null normals (for which $\kappa = 0$), the null surface term in eq.~\eqref{surface} vanishes. This leaves only the joint terms at the meeting points, $r=r_m^1$ and $r_m^2$. The final result for these joint contributions is given by \cite{Growth}
\begin{equation}\label{RNjoints}
I_{\text{joint}} = - \frac{1}{2 \, G_N} \, \left[ (r_m^1)^{2} \log \left[ \frac{|f_{\RNo} (r_m^1) |}{\xi^2} \right] +  (r_m^2)^{2} \log \left[ \frac{|f_{\RNo} (r_m^2) |}{\xi^2} \right]  \right] \, ,
\end{equation}
where $\xi$ is the normalization constant appearing in the null normals, \ie $k\cdot\partial_t|_{r\to\infty}=\pm\xi$. In a moment, the addition of the counterterm \eqref{Counterterm} will eliminate the $\xi$ dependence of the action. Using eq.~\eqref{MeetPntRN}, the time derivative of eq.~\eqref{RNjoints} becomes
\begin{equation}
\frac{d I_{\text{joint}}}{d t} = - \frac{1}{4 G_N} \left[ 2 r f_{\RNo} (r) \log \frac{|f_{\RNo} (r)|}{\xi^2}  + r^{2} \partial_r f_{\RNo} (r) \right]^{r_m^1}_{r_m^2} \, . \label{JnttDer}
\end{equation}
Note that at late times, $r^{1,2}_m$ approach the horizons and so the first term above vanishes. Hence only the second term contributes to the late-time growth rate.

\subsubsection*{Counterterm contribution}

The boundary counterterm~\eqref{Counterterm} requires evaluating the
expansion scalar $\Theta = \partial_{\lambda} \log \sqrt{\gamma}$ in the null boundaries of the WDW patch and the final result is given by
\begin{align}
&I_{\text{ct}}  = \frac{ r_{\text{max}}^{2}}{G_N} \left[ \log \left( \frac{4 \xi^2  \ctL^2}{r_{\text{max}}^{2}} \right) + 1\right] \label{TotalCT} \\
&- \frac{\,( r_{m}^{1})^{2}}{2 G_N}  \, \left[ \log \left( \frac{4 \xi^2  \ctL^2}{( r_{m}^{1})^{2}} \right) + 1  \right] - \frac{ ( r_{m}^{2})^{2}}{2 G_N}  \, \left[ \log \left( \frac{ 4 \xi^2  \ctL^2}{( r_{m}^{2})^{2}} \right) + 1 \right] \, . \nonumber
\end{align}
The term in the first line comes from the UV regulator surface and again only contributes a fixed constant. Hence the time dependence comes only from the terms evaluated at the meeting points in the second line. The time derivative of eq.~\eqref{TotalCT} has a compact form,
\begin{equation}
\frac{d I_{\text{ct}} }{d t} =-\left[  \frac{ r f_{\RNo} (r)}{2  G_N} \,  \log \left( \frac{4 \xi^2 \ctL^2 }{r^2} \right) \right]^{r_m^1}_{r_m^2} \, .
\end{equation}\label{CTtDer}
Again at late times, this contribution vanishes and so it only changes the transient behaviour in the growth rate at early times. It is useful to combine eqs.~\eqref{JnttDer} and \eqref{CTtDer} to explicitly see that the $\xi$ dependence is eliminated, \small
\begin{align}
&\frac{d}{d t} \left( I_{\text{joint}} + I_{\text{ct}} \right) =  - \frac{1}{4 G_N} \left[ 2 r f_{\RNo} (r) \log \left[ \frac{|f_{\RNo} (r)| 4 \ctL^2} {r^2} \right]  + r^{2} \partial_r f_{\RNo} (r) \right]^{r_m^1}_{r_m^2} \, \nonumber \\
&=   - \frac{1 }{4 G_N} \left[ 2 r f_{\RNo} (r) \log \left[ \frac{|f_{\RNo} (r)| 4 \ctL^2} {r^2} \right]  + 2 \frac{r^3}{L^2} - \frac{2  (q_e^2 + q_m^2)}{r} \right]^{r_m^1}_{r_m^2}    \, . \label{JntandCTtDer}
\end{align}
\normalsize
Note that in contrast to eq.~\reef{BulktDer}, the electric and magnetic charges contribute with the same sign above.

\subsubsection*{Total growth rate}

The growth rate of the holographic complexity \reef{defineCA} is then given by the sum of eqs.~\eqref{BulktDer} and \eqref{JntandCTtDer}, which yields
\begin{equation}
\frac{d \mathcal{C}_A}{d t} = \frac{1}{\pi} \frac{d}{d t} \left(  I_{\text{bulk}}  + I_{\text{joint}} + I_{\text{ct}}   \right) =  \frac{q_e^2}{\pi G_N r} \bigg{|}^{r_m^1}_{r_m^2}  - \frac{r \, f_{\RNo} (r) }{2 \pi G_N} \log \left[ \frac{|f_{\RNo} (r)| 4 \ctL^2} {r^2} \right]^{r_m^1}_{r_m^2}   \, . \label{CARNdyo}
\end{equation}
At late times, the past and future meeting points meet the outer and inner horizons, respectively, and so the second term vanishes (since $f_{\RNo} (r_\pm)=0$). This leaves the surprising result 
\begin{equation}
 \lim_{t \to \infty}\frac{d \mathcal{C}_A}{d t}   =  \frac{q_e^2}{\pi G_N\, r} \bigg{|}^{r_-}_{r_+}  \, . \label{square}
\end{equation}
Hence if we consider a purely magnetic black hole with $q_e = 0$, the growth rate vanishes!  More generally, we might introduce 
\begin{equation}\label{brick}
 \qt^2\equiv q_e^2 + q_m^2    \qquad \text{and} \qquad \, \chi \equiv \frac{q_e}{q_m}  \, ,
\end{equation}
which allows us to re-express eq.~\reef{square} as
\begin{equation}
 \lim_{t \to \infty}\frac{d \mathcal{C}_A}{d t}   = \frac{\chi^2}{1+\chi^2}\, \frac{\qt^2}{\pi G_N\, r} \bigg{|}^{r_-}_{r_+}  \, . \label{squareB}
\end{equation}
Now fixing $\qt$, which fixes the spacetime geometry (\eg $r_\pm$), this expression reveals a nontrivial dependence of this growth rate on $\chi$, the ratio of the electric and magnetic charges. In particular, we see that as we put more of the charge $\qt$ into the magnetic monopole with $\chi\to0$, the late-time growth rate shrinks to zero.

\begin{figure}
\centering
\includegraphics[scale=1]{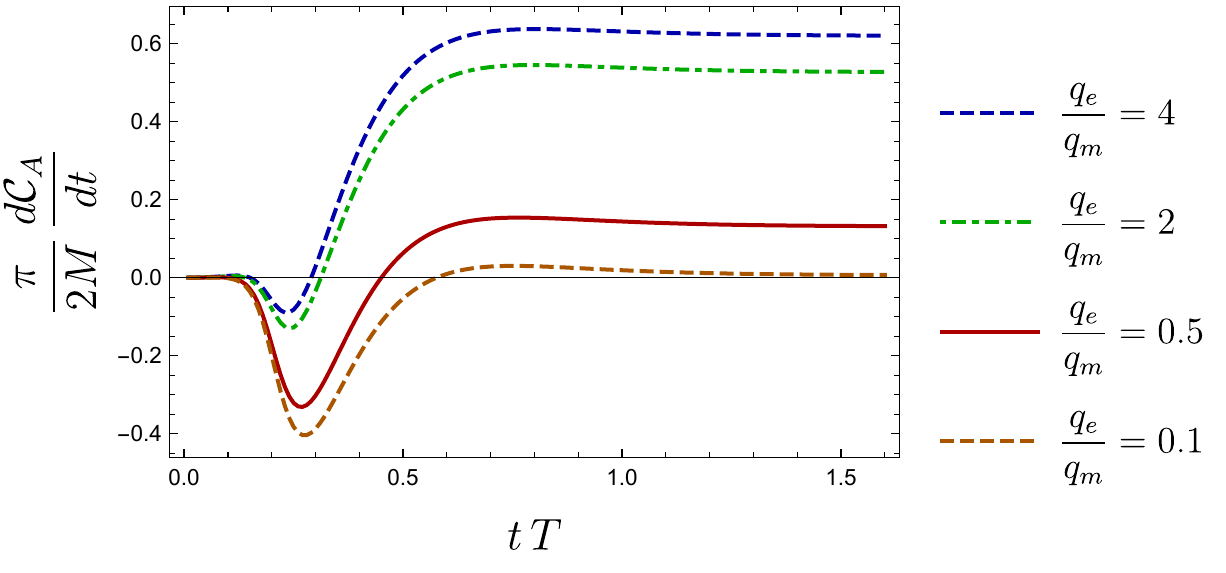}
\caption{The rate of change of complexity for the dyonic black hole given by eq.~\eqref{RNBas}, with $r_-=0.3\,r_+$, $L=0.5\,r_+$ and $\ctL = L$. We fix the parameters that determine the geometry, but vary the ratio between electric and magnetic charges. As predicted by eq.~\eqref{square}, when the charge is mostly magnetic, the growth rate of complexity approaches zero at late times. The limit $q_m \rightarrow 0$  essentially matches the top curve for $\chi=10$. Similarly the $q_e \rightarrow 0$ and the $\chi=0.1$ curves are indistinguishable on this scale. }
\label{Fig:DyonicEter}
\end{figure}

Figure \ref{Fig:DyonicEter} illustrates the full time-dependence of the growth rate, as we change the ratio of the electric and magnetic charges while keeping the spacetime geometry fixed. As expected from eq.~\eqref{square}, the rate approaches zero at late times when the black hole is mostly magnetic. 

\subsection{Maxwell Boundary Term}\label{sec:MBA}

The discussion in the previous section raises the question of whether there is a consistent prescription for the holographic complexity that puts the electric and magnetic charges on an equal footing? In the following, we will argue that such a prescription requires that we modify the action with the addition of the Maxwell boundary term in eq.~\reef{mba0} 
\begin{equation}\label{mba2}
I_{\mu\mt{Q}} = \frac{\ap}{g^2} \int_{\partial \mathcal{M}} d \Sigma_{\mu} \, F^{\mu \nu} \, A_{\nu} \, .
\end{equation}
This surface term plays a natural role in black hole thermodynamics \cite{char0} --- see \cite{char1,char2} for a discussion in the context of the AdS/CFT correspondence. In particular, the Euclidean version of the action $I_0$ would yield the Gibbs free energy, associated with the grand canonical ensemble where the temperature and chemical potential $\mu$ are held fixed. Adding the boundary term \reef{mba2} (with $\ap=1$) to the Euclidean action produces the Legendre transform to the Helmholtz free energy, associated with the canonical ensemble where the temperature and total (electric) charge $Q$ are held fixed. This boundary term was also shown to play a role in resolving the apparent tension between electric-magnetic duality in four dimensions and  the different partition functions of electric and magnetic black holes \cite{Hawking:1995ap, Deser:1996xu, Brown:1997dm} --- see further discussion below. 

As we noted above, adding this surface term \reef{mba2} changes the boundary conditions in the variational principle of the Maxwell field. Consider varying the Maxwell action in eq.~\reef{bulk}. Integrating by parts produces the equations of motion in the bulk but leaves a boundary term proportional to $\delta A_{\mu}$,
\begin{equation}
\delta I_{\mt{Max}}  =  \frac{1}{ g^2} \int_{\mathcal{M}} d^{4} x \sqrt{-g} \,  \nabla_\mu F^{\mu \nu} \, \delta A_{\nu} - \frac{1}{g^2} \, \int_{\partial \mathcal{M}}  d \Sigma_{\mu} \, F^{\mu \nu} \, \delta A_{\nu} \, .
\end{equation}
Hence, a well-posed variational principle requires a Dirichlet boundary condition setting $\delta A_a =0$ on the boundary (where the index $a$ indicates that only the tangential components of the potential are fixed). However, the latter can be modified by introducing the surface term \reef{mba2}, in which case the variation produces the boundary contribution
\begin{equation}
\delta I_{\mt{Max}}  +\delta I_{\mu\mt{Q}}= \cdots  - \frac{1}{g^2} \, \int_{\partial \mathcal{M}}  d \Sigma_{\mu} \left[(1-\ap)\, F^{\mu \nu} \, \delta A_{\nu}+ \ap \,\delta  F^{\mu \nu} \,  A_{\nu} \right]\, .
\end{equation}
Of course, with $\ap=1$, the term proportional to $\delta A_\nu$ is eliminated and the required boundary condition becomes $ n^\mu \,\delta F_{\mu a} =0 $, where $n^\mu$ is a unit vector orthogonal to the boundary $\partial \mathcal{M}$. If we choose a gauge where $n\cdot A=0$, we recognize this as the Neumann boundary condition $ n^\mu \,\partial_\mu\, \delta A_a =0$. With a general value of $\ap$ (and the same choice of gauge), the potential would satisfy a mixed boundary condition,
\beq
\ap\,n^\mu\partial_\mu \delta A_a =( 1-\ap)\,X_a{}^b\,\delta A_b\,,
\label{mixed}
\eeq
where the choice of $X_a{}^b$ will depend on details of the problem of interest, \eg \cite{Vassilevich:2003xt, Esposito:2004ts, robin}.

Returning to the action \reef{TOTaction}, if we use the Maxwell equations $\nabla_\mu F^{\mu\nu}=0$, then the boundary term \eqref{mba2} can be converted into a bulk term via Stokes' theorem as
\begin{equation}\label{MaxBdryActOnS}
I_{\mu\mt{Q}}\big{|}_{\mt{on-shell}}  = \frac{\ap}{2g^2}  \int_{\mathcal{M}} d^{4} x \sqrt{-g}\, F_{\mu \nu} F^{ \mu \nu} \, ,
\end{equation}
which is explicitly gauge invariant.\footnote{There is a subtlety here for the magnetic monopole contribution in that the boundary term must be integrated over the boundary of all patches where the potential is well-defined --- see appendix \ref{MBAapp}.} Of course, the above expression takes the same form as the bulk Maxwell action \reef{bulk} and so we could just as well have re-expressed the bulk action as a boundary term. In any event, combining eq.~\reef{MaxBdryActOnS} with $I_\mt{Max}$ yields
\beq\label{combo}
I_\mt{Max}+I_{\mu\mt{Q}}\big{|}_{\mt{on-shell}}  = \frac{2\ap-1}{4g^2}  \int_{\mathcal{M}} d^{4} x \sqrt{-g}\, F_{\mu \nu} F^{ \mu \nu} \, .
\eeq
Hence in evaluating the WDW action for the general action $I_\mt{tot}(\ap)$, \ie including the contribution of the Maxwell boundary term in eq.~\reef{TOTaction}, the only change that has to be made to the previous calculation is to change the overall coefficient of the Maxwell contribution in eq.~\reef{BulkIntegrand}. As a result, eq.~\reef{BulktDer} is replaced by
\begin{equation}
\frac{d \ }{d t}\left(I_{\text{bulk}} +I_{\mu\mt{Q}} \right) =   \frac{1}{2 G_N} \left[ \frac{r^3}{L^2} -(2\ap -1)\,\frac{q_e^2- q_m^2}{r} \right]_{r_m^2}^{r_m^1} \, . \label{BtDer2}
\end{equation}

Subsequently, the final result for the late-time growth rate for the complexity becomes
\begin{equation}
 \lim_{t \to \infty}\frac{d \mathcal{C}_A}{d t}   =  \frac{ (1-\ap)\,q_e^2+ \ap\,q_m^2}{\pi G_N r} \bigg{|}^{r_-}_{r_+} =\frac{(1-\ap)\chi^2+\ap}{1+\chi^2}\, \frac{\qt^2}{\pi G_N r}\bigg{|}^{r_-}_{r_+}  \, . \label{square2}
\end{equation}
Therefore if we set $\ap=1$, the dependence on the electric charge drops out of the numerator and the late-time growth rate is primarily sensitive to the magnetic charge. In particular then, with this choice of $\ap$, the late-time growth rate drops to zero for an electrically charged black hole at late times. 

The above discussion shows us that the growth rate (or more generally the on-shell action) is symmetric under electric-magnetic duality, \ie $F_{\mu\nu}\leftrightarrow
\widetilde F_{\mu\nu}=\frac12 \varepsilon_{\mu\nu\rho\sigma} F^{\rho\sigma}$, if at the same time we exchange the action\footnote{This equivalence was noted by \cite{Brown:1997dm} for $\gamma=1$.} 
\beq
I_\mt{tot}(\ap)\leftrightarrow
I_\mt{tot}(1-\ap)\,,
\label{exchange}
\eeq
\ie we modify the coefficient of the Maxwell boundary term \reef{mba2} as indicated above. Then $\ap=1/2$ is singled out as the special choice which leaves the action unchanged in eq.~\reef{exchange}. Of course, looking back at eq.~\reef{combo}, we see that the combination of the bulk and boundary terms for the Maxwell field vanishes on-shell. However, the complexity is still sensitive to the electromagnetic field through its back-reaction on the geometry. In particular, the holographic complexity only depends on the duality invariant combination $\qt^2=q_e^2+q_m^2$, as appears in the metric \reef{RNBas}. For example, eq.~\reef{square2} becomes 
\begin{equation}
 \lim_{t \to \infty}\frac{d \mathcal{C}_A}{d t}  
 \bigg|_{\ap=1/2} =  \frac{q_e^2+ q_m^2}{2\pi G_N\, r} \bigg{|}^{r_-}_{r_+}  \, , \label{square3}
\end{equation}
and as desired, the electric and magnetic charges influence the complexity growth rate on an equal footing. However, as we discuss in section \ref{sec:Discussion}, this expression produces a puzzle in the limit of zero charges.

Of course, the reader may wonder why we should expect that 
that magnetic and electric black holes should compute at the same rate. First, let us recall the expectation that the late-time growth of the complexity should be given by eq.~\reef{general}, \ie $d{\cal C}/dt \sim ST$, but both the entropy $S$ and temperature $T$ are governed by the spacetime geometry, as given in eq.~\reef{MST}. Hence it is natural to think that this rate should be controlled by $\qt^2=q_e^2+q_m^2$, the combination appearing in the metric \reef{RNBas}. This conclusion can also be motivated by the shock wave geometries, which we study in the next section. In this context, both electric or magnetic black holes exhibit the same back-reaction and hence it is natural to think that the holographic complexity should respond in the same manner independent of the nature of the charge.

\subsection{Shock Wave Geometries}\label{subsec:Shocks}
 
Another property that holographic complexity should exhibit is the {\it switchback effect}, which is related to the complexity of precursor operators \cite{Stanford:2014jda, Bridges} --- see further discussion in section \ref{sec:Discussion}. We will follow closely the analysis and notation of \cite{Vad1,Vad2}.
To examine this feature, we consider a Vaidya geometry where a(n infinitely) thin shell of null fluid collapses into a charged black hole. If the shell only injects a small amount of energy into the system,  then the black hole's event horizon shifts by a small amount, \ie
\begin{equation}
\frac{r_{+,2}}{r_{+,1}} = 1 + \epsilon \, ,
\label{house}
\end{equation}
where the subscripts 1 and 2 indicate before and after the shock wave, respectively.
The scrambling time associated with this perturbation is then given by
\begin{equation}\label{Scramblingt}
t^{*}_{\text{scr}} = \frac{1}{2 \pi T_1} \log \frac{2}{\epsilon} \, .
\end{equation}
For the chaotic dual of the black hole, the  switchback effect then predicts that for any time $t$ after the perturbation is introduced, the complexity remains essentially unchanged for  $t<t^{*}_{\text{scr}}$ but then the difference of complexities (for the perturbed and unperturbed states) begins to grow linearly  afterwards, \ie $t>t^{*}_{\text{scr}}$. Our goal here is to investigate to what extent the CA proposal reproduces this behaviour for the charged black holes discussed in the previous sections.

\subsubsection*{Charged shock wave geometry}  

Figure \ref{PenroseHigherCharged}(b) illustrates the spacetime geometry for a shock wave collapsing into a Reissner-Nordstrom black hole from the right boundary at $t=-t_w$. Note that following \cite{Vad1,Vad2}, we adopt the Dray-`t Hooft prescription that the null geodesics shift upon crossing the collapsing shock wave. For simplicity, we assume that the thin shell is neutral, \ie it carries energy but no charges.
The corresponding metric is 
\begin{align}
& d s^2 = -F(r,v)\, d v^2 + 2\, d r\, d v + r^2\, (d \theta^2 + \sin ^2 \theta d \phi^2 )  \nonumber \\
&\quad{\rm with} \qquad F(r, v)= \frac{r^2}{L^2} + 1 - \frac{f_1(v)}{r} + \frac{q_e^2+q_m^2}{r^{2}}  \label{shocker}
\end{align}
where
\begin{equation}
f_s (v) = \omega_1\, (1 - \mathcal{H}(v-v_s)) + \omega_2\,  \mathcal{H}(v-v_s) \, . \nonumber \\
\end{equation}
(with $\mathcal{H}(v)$ denoting the usual Heaviside function). Before and after $v=v_s$, the metric has precisely the form given in eq.~\reef{RNBas} with $\omega=\omega_1$ and $\omega_2$, respectively. However, we must evaluate the tortoise coordinate \reef{rstar} for each region and then following eq.~\reef{uv}, define the time coordinate as $t=v-r^*(r)$. Note that taking the limit $r\to\infty$, we find $v_s=-t_w$ on the boundary.

The geometry of the WDW patch is characterized by a number of dynamical points: $r_m^1$ and $r_m^2$, the meeting points of the future and past null boundaries, respectively; and $r_s$ and $r_b$, the point where the null shell crosses the past right and future left boundaries, respectively. These positions are determined by the boundary times with
\begin{align}
&t_R + t_w = - 2 r^*_2 (r_s) \, , \nonumber \\
&t_L - t_w = 2 r^*_1 (r_s) - 2 r^*_1 (r_m^2) \, ,\nonumber \\
&t_L - t_w = 2 r_1^* (r_b) \, , \nonumber \\
& t_R + t_w = 2 r^*_2 (r_m^1) - 2 r^*_2 (r_b)\, . \label{eq:ShoPos}
\end{align}
In the following, it is sufficient to restrict our attention to the case $t_L=t_R = 0$ and to study the behaviour resulting from pushing the perturbation to earlier times $t=-t_w$. Let us note that with these choices, eq.~\eqref{eq:ShoPos} yields a simple result for the dynamical points in the limit of large $t_w$, namely,
\begin{align}
& \lim_{t_w T \rightarrow \infty} r_s = r_{+, 2} \, \qquad \qquad \,  \lim_{t_w T \rightarrow \infty} r_m^2 = r_{-, 1} \nonumber \\
&  \lim_{t_w T \rightarrow \infty} r_b = r_{+, 1}  \, \qquad \qquad \,   \lim_{t_w T \rightarrow \infty}  r_m^1 = r_{-, 2}  \, .
\end{align}

\subsubsection*{Results for Switchback Effect}

Following \cite{Vad2}, the switchback effect is revealed (or not) in the `complexity of formation' comparing the holographic complexity of the above shockwave geometry with  that of the static black hole \reef{RNBas} with $\omega=\omega_1$ (and the same charges). We begin by considering the CA proposal for the action without the Maxwell boundary term \reef{mba0}, \ie we again set $\ap=0$ in eq.~\reef{TOTaction} as in section \ref{subsec:CompDyon}.  The details of our calculations are given in appendix \ref{App:Shock}. In figure \ref{DyonShockChi}, we present the difference of complexities for a light shock wave producing $r_{+,2} = (1+10^{-6}) r_{+,1}$ (\ie $\epsilon=10^{-6}$ in eq.~\reef{house}). Notice that the complexity remains unchanged by the perturbation up until $t_w=t^{*}_{\text{scr}}$ but afterwards, the difference quickly makes a transition to linear growth.\footnote{For heavier shock waves, \eg $\epsilon\sim 10^{-1}$, the initial regime over which the complexity is constant essentially disappears, similar to the behaviour found for neutral black holes in \cite{Vad2}.} Several curves are shown in the figure where the geometry is held fixed (\ie $\qt^2=q_e^2+q_m^2$ is fixed) but the ratio $\chi=q_e/q_m$ is varied.  We see that the rate of the linear growth $t_w\ge t^{*}_{\text{scr}}$ decreases to zero as more of the charge is put into the magnetic monopole, \ie as $\chi\to0$. Hence the switchback effect vanishes (with this choice of $\ap$) for a black hole with pure magnetic charge. This result might be expected since there is a close connection between the late-time rate of growth of the complexity in the static black hole and $d{\cal C}_A/dt_w$, as discussed in \cite{Vad2}.\footnote{Comparing eq.~\reef{squareB} with the result in eq.~\reef{VaidyaDyon} for $t_w > t^{*}_{\text{scr}}$, we see that $d{\cal C}_A/dt_w\simeq 2\,d{\cal C}_A/dt|_{t\to\infty}$, as predicted by \cite{Vad2}.}

\begin{figure}
\centering
\includegraphics[scale=1]{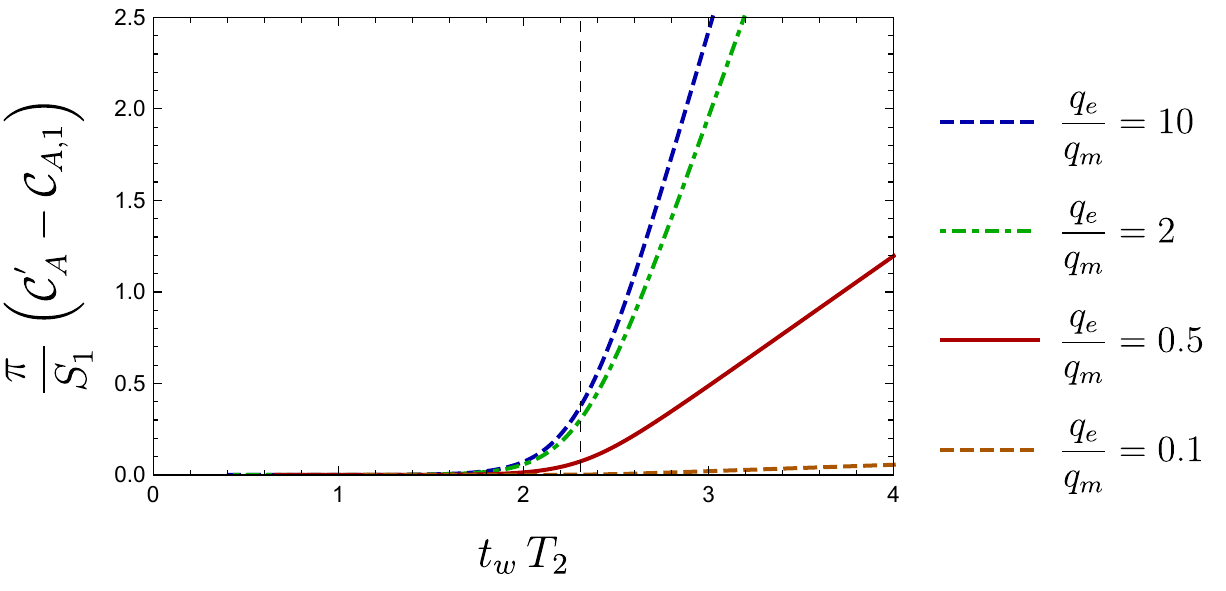}
\caption{The difference of complexities of formation in the shock wave geometry as a function of the insertion time $t_w$, for a light shock wave $r_{+,2} = (1+10^{-6}) r_{+,1}$ and parameters $L= 0.5\, r_{+,2}$, $r_{-,1} = 0.5\, r_{+,2}$ and $\ell_{\text{ct}} =L$ ($r_{-,2}$ is then fixed by the condition that $\qt$ is the same after the shock wave). The dashed vertical line is the scrambling time for the shock wave with these geometric parameters. We investigate the effect of varying the ratio between electric and magnetic charges: after the scrambling time, the complexity essentially remains constant for the solution with mostly magnetic charges, as predicted by eq.~\eqref{VaidyaDyon}. }
\label{DyonShockChi}
\end{figure}

The rate $d \mathcal{C}_A/d t_w$ can be evaluated analytically to find (see appendix \ref{App:Shock})
\begin{align}
&\frac{d \mathcal{C}_A}{d t_w} \simeq\ \mathcal{O} \left(\frac{\chi^2}{1+\chi^2} \,  
\epsilon e^{2 \pi T_1 t_w}
\right) \, \qquad\qquad \text{for} \, t_w < t^{*}_{\text{scr}} \, , \nonumber \\
&\frac{d \mathcal{C}_A}{d t_w} \simeq \frac{\chi^2}{1+\chi^2}\, \frac{\qt^2}{\pi G_N} \left( \frac{ 1}{r}\, \bigg{|}^{r_{-,1}}_{r_{+,1}} + \frac{ 1}{r}\,  \bigg{|}^{r_{-,2}}_{r_{+,2}}  \right)   \, \qquad \text{for} \, t_w > t^{*}_{\text{scr}} \, .
  \label{VaidyaDyon}
\end{align}
Hence as for the growth rate of the eternal black hole case in section \ref{subsec:CompDyon}, the complexity rate after the scrambling time depends on the ratio between electric and magnetic charges and in particular, $d \mathcal{C}_A/d t_w$ vanishes as $\chi\to0$.  

\begin{figure*}[htp]
    \centering
    \begin{subfigure}{0.5\textwidth}
      \centering
\includegraphics[scale=0.75]{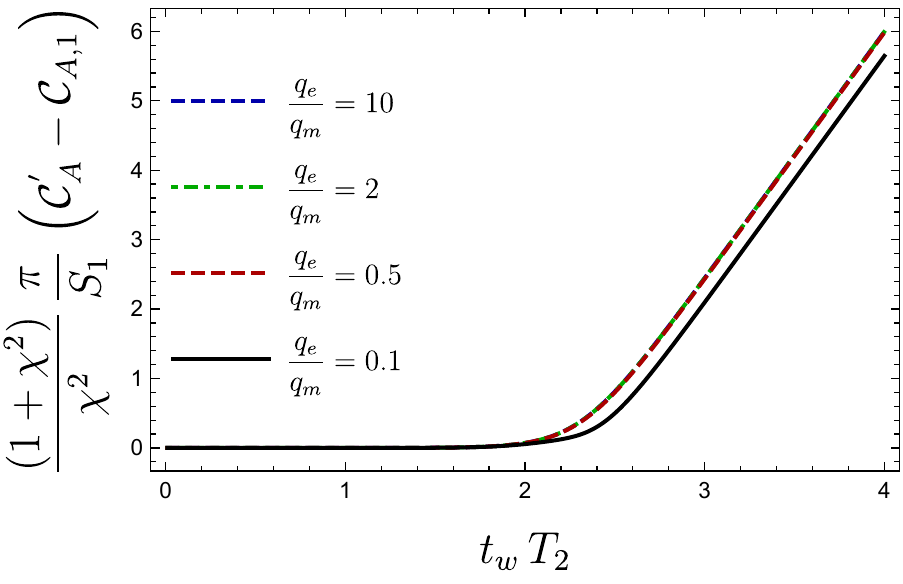}
      \caption{ }
    \end{subfigure}%
    \begin{subfigure}{0.5\textwidth}
      \centering
\includegraphics[scale=0.75]{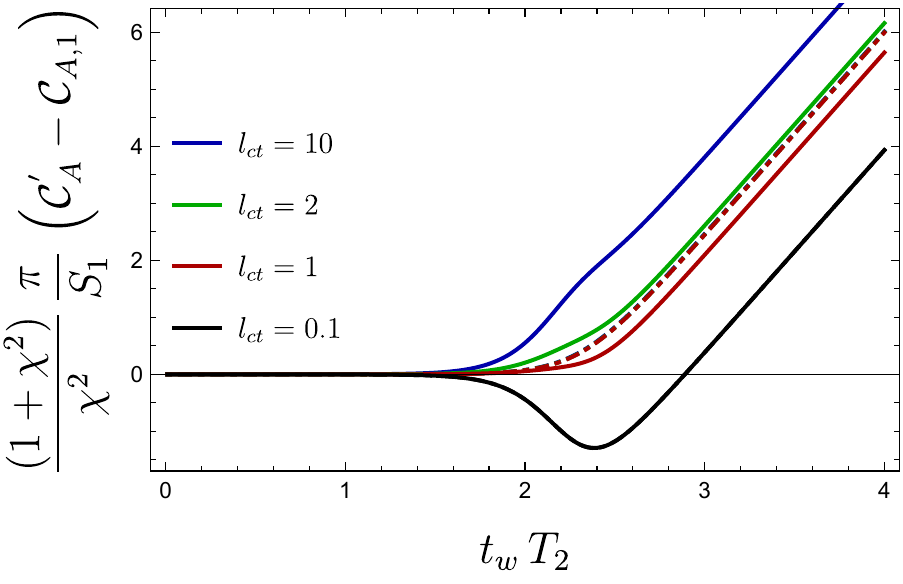}
      \caption{}
    \end{subfigure}
  \caption { (a) The complexity for the shock wave geometry as a function of the insertion time $t_w$, for a light shock wave $r_{+,2} = (1+10^{-6}) r_{+,1}$ and parameters $L= 0.5\, r_{+,2}$, $r_{-,1} = 0.5\, r_{+,2}$ and $\ctL =L$. (Further, $r_{-,2}$ is determined by fixing $\qt$ to be the same before and after the shock wave.) We show the result for rescaling the curves in figure\ref{DyonShockChi}  by the factor $(1+\chi^2)/\chi^2$. We essentially see that the curves lie on top of each other, except the for smallest $\chi$, as it is more sensitive to the transient behaviour controlled by $\ctL$.  (b) The influence of the transient behaviour for the complexity in the shock wave geometry. We show in solid $\chi= 0.1$ and in dot-dashed $\chi=10$ to contrast the effect of varying $\ctL$. For $\ctL =0.1\,L$, both curves are essentially on top of each other, but for $\ctL \sim L$, the curves with small $\chi$ are more sensitive to this ambiguous scale. }
  \label{Fig:res}
\end{figure*}

We can confirm the scaling with $\chi$ in eq.~\eqref{VaidyaDyon} by simply multiplying the curves in figure \ref{DyonShockChi} by the factor $(1+\chi^2)/\chi^2$ and then we see in figure \ref{Fig:res}(a) that essentially they all collapse onto a single curve. The only exception is for the smallest ratio, $\chi=0.1$, which is slightly shifted to the right. This behaviour arises  because for smaller $\chi$, there is a greater sensitivity to the scale  $\ctL$ in the null counterterm \reef{Counterterm}. The dependence on this ambiguity in the definition of the WDW action is illustrated in figure \ref{Fig:res}(b). We note, however, that this ambiguity does not effect the final rate $d \mathcal{C}_A/d t_w$ but only the transition between the two regimes in eq.~\reef{VaidyaDyon}.

\begin{figure}
\centering
\includegraphics[scale=1]{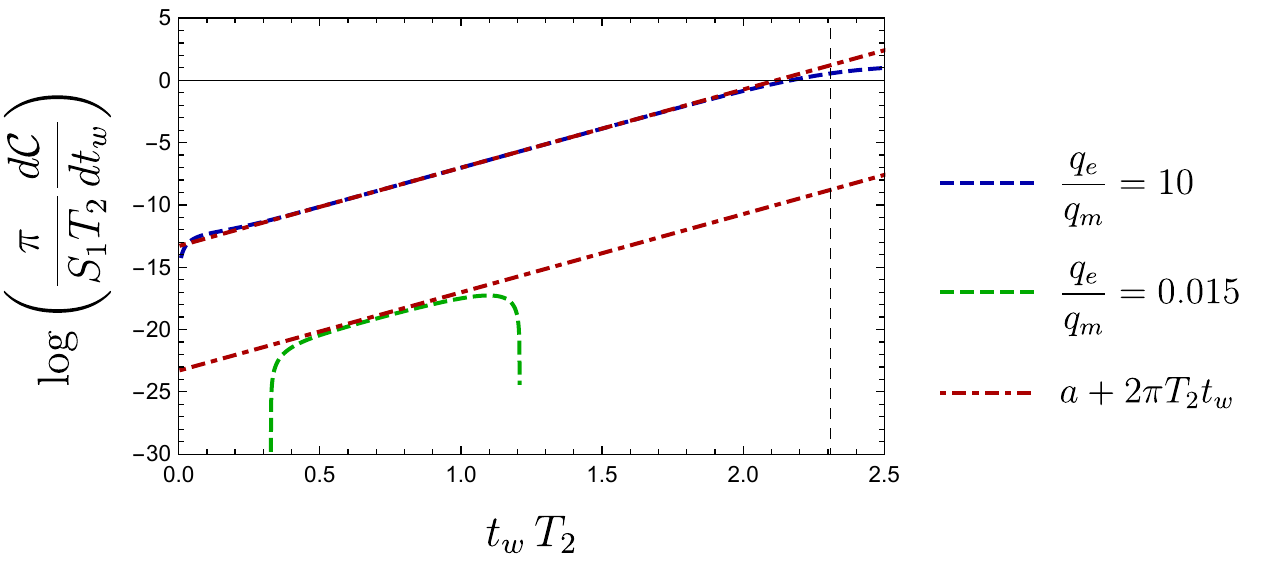}
\caption{The exponential growth for the complexity with a light shock wave, such that $r_{+,2} = (1+10^{-6}) r_{+,1}$, and also with $L= 0.5 r_{+,2}$, $r_{-,1} = 0.5 r_{+,2}$, $l_{\text{ct}} =L$. We show two examples, one for a black hole with mostly electric charge $\chi=10$ and one with mostly magnetic charge $\chi =0.015$. For the larger $\chi$, the dynamics is well approximated by an exponentially growing mode with Lyapunov exponent $\lambda_L = 2 \pi T$ until times of the order of the scrambling time (vertical black line), as in eq.~\eqref{VaidyaDyon}. For the smaller value of $\chi$, the amplitude of this initial mode is suppressed both by the energy of the shock wave and  by a factor of $\chi^2$. The exponentially growing mode only dominates at very early times because it must compete with other transient effects. In the limit that the black hole has only magnetic charges
(\ie $\chi=0$), this exponentially growing mode is absent. }
\label{DyonLyapu}
\end{figure}

Further, eq.~\reef{VaidyaDyon} suggests a regime of exponential growth for $t_w < t^{*}_{\text{scr}}$. However, this regime actually becomes smaller as the black hole becomes mostly magnetic, \ie as $\chi$ becomes small, as illustrated in figure \ref{DyonLyapu}. For the mostly electric black hole (with $\chi=10$), we see a good agreement with an exponentially growing mode with the Lyapunov exponent $\lambda_L = 2 \pi T$ until times of the order of the scrambling time. For the mostly magnetic black hole (with $\chi =0.015$), the amplitude of the exponential mode is suppressed by a factor of $\chi^2$, which shifts the corresponding curve down in the figure. In addition, we see that the exponentially growing mode is only the dominant contribution at earlier times. This reflects the fact that the analysis producing the $t_w < t^{*}_{\text{scr}}$ expression in eq.~\reef{VaidyaDyon} really only applies for $\chi \gtrsim 1$ --- see appendix \ref{App:Shock}. When this exponential mode is suppressed by small $\chi$, it must compete with other transient dynamics (\eg depending on $\ell_{\text{ct}}$) and therefore, its role becomes less important in this regime. In particular, if the black hole is purely magnetic (with $\chi=0$), the exponentially growing mode is absent.  

Of course, the above results (with $\ap=0$) are modified if we include the Maxwell boundary term \reef{mba0}. In particular, eq.~\reef{VaidyaDyon} is replaced by
\begin{align}
&\frac{d \mathcal{C}_A}{d t_w} \simeq\ \mathcal{O} \left(\frac{(1-\ap)\chi^2+\ap}{1+\chi^2} \,  
\epsilon e^{2 \pi T_1 t_w}
\right) \, \qquad\qquad \text{for} \, t_w < t^{*}_{\text{scr}} \, , \nonumber \\
&\frac{d \mathcal{C}_A}{d t_w} \simeq \frac{(1-\ap)\chi^2+\ap}{1+\chi^2}\, \frac{\qt^2}{\pi G_N} \left( \frac{ 1}{r}\, \bigg{|}^{r_{-,1}}_{r_{+,1}} + \frac{ 1}{r}\,  \bigg{|}^{r_{-,2}}_{r_{+,2}}  \right)   \, \qquad \text{for} \, t_w > t^{*}_{\text{scr}} \, .
  \label{VDon}
\end{align}
Hence if we choose $\ap=1$, the roles of the magnetic and electric charges are reversed. For example, with this choice, black holes with magnetic charges exhibit the desired switchback effect while those with a purely electric charge would not. Further, similar to the discussion in section \ref{sec:MBA}, if we choose $\ap=1/2$, the $\chi$ dependence drops out of eq.~\reef{VDon} and the behaviour only depends on $\qt^2=q_e^2+q_m^2$. Therefore, with this choice, both electric and magnetic black holes exhibit the same switchback effect.

As a final comment here, let us note that for a light shock wave, $r_{\pm,2}\simeq r_{\pm,1}$ and the two contributions in eq.~\reef{VDon} for the rate at large $t_w$ are essentially the same. Further, this rate is essentially twice the late-time growth rate in eq.~\reef{square2}.\footnote{Of course, there is a similar relationship between the rates in eqs.~\reef{VaidyaDyon} and \reef{squareB} for $\gamma=0$.} In fact, as discussed in \cite{Vad2}, there is a more general relationship that extends to heavy shocks, \ie
\beq
\frac{d \mathcal{C}_A}{d t_w}=\frac{d \mathcal{C}_A}{d t_\mt{R}}-\frac{d \mathcal{C}_A}{d t_\mt{L}}\,,
\label{relate}
\eeq
because of the symmetry of the shock wave geometry under
\beq
t_\mt{R}\to t_\mt{R}-\Delta t\,,\qquad
t_\mt{L}\to t_\mt{L}+\Delta t\,,\qquad
t_w\to t_w +\Delta t\,.
\label{shifter}
\eeq
Hence for large $t_w$, $d \mathcal{C}_A/d t_w$ is related to the late-time growth rates of the complexity on either side of the shock wave.\footnote{That is, given a large $t_w=t_0$, we use eq.~\reef{shifter} to shift
$(t_\mt{R},t_\mt{L},t_w)=(0,0,t_0)\to(t_0,-t_0,0)$. Then the right-hand side of eq.~\reef{relate} has a contribution corresponding to the growth rate on the right boundary at very late times and another coming from very early times on the left boundary. In fact, the latter is probing the white hole part of the Penrose diagram when the complexity is actually decreasing, \eg \cite{Susskind:2015toa}. However, by the time symmetry of the unperturbed Penrose diagram, this early-time rate matches the late-time rate up to an overall sign, \ie the minus sign in eq.~\reef{relate}. } Therefore, we can anticipate that the switchback effect will be absent in exactly the same situations where the late-time growth rate vanishes, \eg for magnetic black holes with $\gamma=0$. Of course, this is precisely the behaviour found in this subsection.

\section{Charged Dilatonic Black Hole}\label{sec:Dila}
In this section, we investigate the CA proposal \reef{defineCA} in a broader class of charged black holes. This investigation is motivated by the question of understanding to what extent our results in the previous section are special to the precise couplings of the Einstein-Maxwell theory. In particular, in many string theoretic settings, the gauge field will also be coupled to various moduli or scalars,  \eg see \cite{Youm:1997hw,Maeda:2011sh}. The presence of these new couplings lead to scalar hair on the charged black holes, and may change the nature of the spacetime singularities and the casual structure of the corresponding black holes. Hence we would like to understand if these changes to the spacetime geometry modify the behaviour of the holographic complexity in an essential way.

 In the following, we consider a simple extension of the Einstein-Maxwell theory, where Maxwell field has an ``exponential coupling" to an additional scalar field, the so-called dilaton.  The corresponding charged dilatonic black holes were introduced for asymptotically flat geometries in \cite{Gibbons:1987ps,Garfinkle:1990qj} and they were extended to asymptotically AdS geometries in \cite{Gao:2004tu,Gao:2004tv,Elvang:2007ba}. The  AdS solutions were further explored in, \eg \cite{Charmousis:2010zz,Gouteraux:2011ce,Hendi:2010gq, Ong:2012yf}.  Holographic complexity of dilatonic black holes has been previously studied for several models \cite{brian,Cai:2017sjv,An:2018xhv,Mahapatra:2018gig}. Our investigation of the holographic complexity for these dilatonic black holes will show that the vanishing of the late-time growth rate found in the previous section (for certain choices of charges and boundary terms) is not a generic result for charged black holes. Rather, for the theories studied here, the analog of the Maxwell boundary term modifies the complexity growth rate but the coefficient can not be chosen to reduce the rate to zero generally. However, we will see that the latter can still be accomplished in the theories where the charged black holes have the same causal structure as the Reissner-Nordstrom black holes. Hence our conclusion is that the causal structure of the spacetime geometry is the essential feature leading to the vanishing late-time growth rate in the previous section.

As commented above, we will be studying holographic complexity in a theory where gravity couples to a dilaton, as well as the Maxwell field (and cosmological constant),
\begin{equation}\label{pda}
I_\mt{bulk} = \ \frac{1}{16 \pi G_N} \int_\mathcal{M} d^{4} x \sqrt{-g} \left(\mathcal{R}-2(\partial \phi )^2-V(\phi)\right)-\frac{1}{4g^2}\int_{\mathcal{M}} d^{4}x \sqrt{-g}e^{-2\alpha\phi}F_{\mu \nu}F^{\mu \nu}\,
\end{equation}
where the dilaton potential $V(\phi)$ given by
\begin{equation}
V(\phi)=-\frac{2}{(1+\alpha^2)^2L^2}\left[\alpha^2(3\alpha^2-1) e^{-2\phi/\alpha}+(3-\alpha^2)e^{2\alpha\phi}+8\alpha^2e^{( \alpha-1/\alpha)\phi} \right]\,.\label{potter}
\end{equation} 
The total action takes the form
\begin{equation}\label{TOTDilaton}
I_\mt{tot}=I_\mt{bulk}+I_\mt{surf}+I_\mt{ct}\,,
\end{equation}
where the gravitational boundary terms, $I_\mt{surf}$ and $I_\mt{ct}$, are the same as in eqs.~\reef{surface} and \reef{Counterterm}, respectively. In subsection \ref{blt2}, we will also consider the effect of adding the analog of the Maxwell surface term \reef{mba0}, as well as a new boundary term for the dilaton. Here, we are again focusing on the case of four bulk dimensions for simplicity.

The parameter $\alpha$ controls the strength of the coupling of the dilaton to the Maxwell field, but it also determines the shape of the potential in eq.~\reef{potter}.  The latter is tuned so that $\phi=0$ is a critical point (\ie a local maximum) with $V(0)=-6/L^2$, where $L$ is the curvature scale of the corresponding AdS vacuum.  We also note that the global shape of the potential depends on the value of $\alpha$, namely,
\begin{itemize}
\item For $0<\alpha^2<1/3$, as well as the maximum at $\phi=0$, ${V}(\phi)$ has a minimum at $\phi=-\frac{\alpha}{1+\alpha^2}\log\left(\frac{1-3\alpha^2}{3-\alpha^2}\right)$. Moreover, $\lim_{\alpha\phi \to \pm \infty} V(\phi)=\mp \infty$.
\item For $1/3 < \alpha^2 < 3$, $\tilde{V}(\phi)$ has only the global maximum at $\phi=0$. In this case, $\lim_{\phi \to \pm \infty} {V}(\phi)=- \infty$.
\item For $\alpha^2>3$, ${V}(\phi)$ has the maximum at $\phi=0$ and a minimum at $\phi=\frac{\alpha}{1+\alpha^2}\log\left(\frac{3\alpha^2-1}{\alpha^2-3}\right)$. Asymptotically, we find $\lim_{\alpha\phi \to \pm \infty} V(\phi)=\pm \infty$.
\item For the special values $\alpha^2=1/3\,,1\,\, \text{and}\,\, 3$,\,\, ${V}(\phi)$ has only a maximum at $\phi=0$, but it is symmetric under $\phi\to-\phi$. More generally, the potential is invariant with the following substitutions: $\phi \to -\phi$ and $\alpha \to {1}/{\alpha}$.
\end{itemize}
Of course, if we set $\al=0$, the dilaton decouples from the Maxwell field and the potential \reef{potter} reduces to a simple cosmological constant, \ie $V(\phi)|_{\alpha=0}=-6/L^2$. Hence in this limit, the theory \reef{pda} reduces to the Einstein-Maxwell theory \reef{bulk} from the previous section coupled to an additional massless scalar field.

For this theory \reef{pda}, a class of static spherically-symmetric solutions describing electrically charged dilaton black holes is given by \cite{Gao:2004tu} 
\beqa\label{metric_dilaton}
&&ds^2=-f(r)\,dt^2+\frac{dr^2}{f(r)}+U^2(r)\,(d\theta^2 +\sin^2\theta\,d\phi^2)\,,\\
&&\quad F=\frac{g}{\sqrt{4\pi G_N}}\frac{q_e\, e^{2\alpha \phi}}{U(r)^2}\, dr \wedge dt\,,\qquad e^{\alpha \phi}=\frac{U(r)}{r}\,,
\nonumber
\eeqa
with
\begin{eqnarray}
&&\quad f(r)=\left(1-\frac{c}{r}\right)\left(1-\frac{b}{r}\right)^{\frac{1-\alpha^2}{1+\alpha^2}}+\frac{U^2(r)}{L^2}\,,
\label{func}\\
&&
U^2(r)=r^2\left(1-\frac{b}{r}\right)^{\frac{2\alpha^2}{1+\alpha^2}},
\qquad
q_e^2=\frac{c\, b}{1+\alpha^2}\,,\nonumber
\end{eqnarray}
where $c$ and $b$ are integration constants.
We note that this solution interpolates between the Reissner-Nordstrom black hole ($\alpha \to 0$) and Schwarzschild ($\alpha \to \infty$).\footnote{In the latter case,  the coordinate transformation $r \to r+b$ yields the usual coordinate system for the Schwarzschild-AdS metric.} Moreover, if we set $b=0$, the solution reduces to the (uncharged) Schwarzschild-AdS solution independently of the value of $\alpha$. 

\begin{figure}
\centering
\includegraphics[scale=0.35]{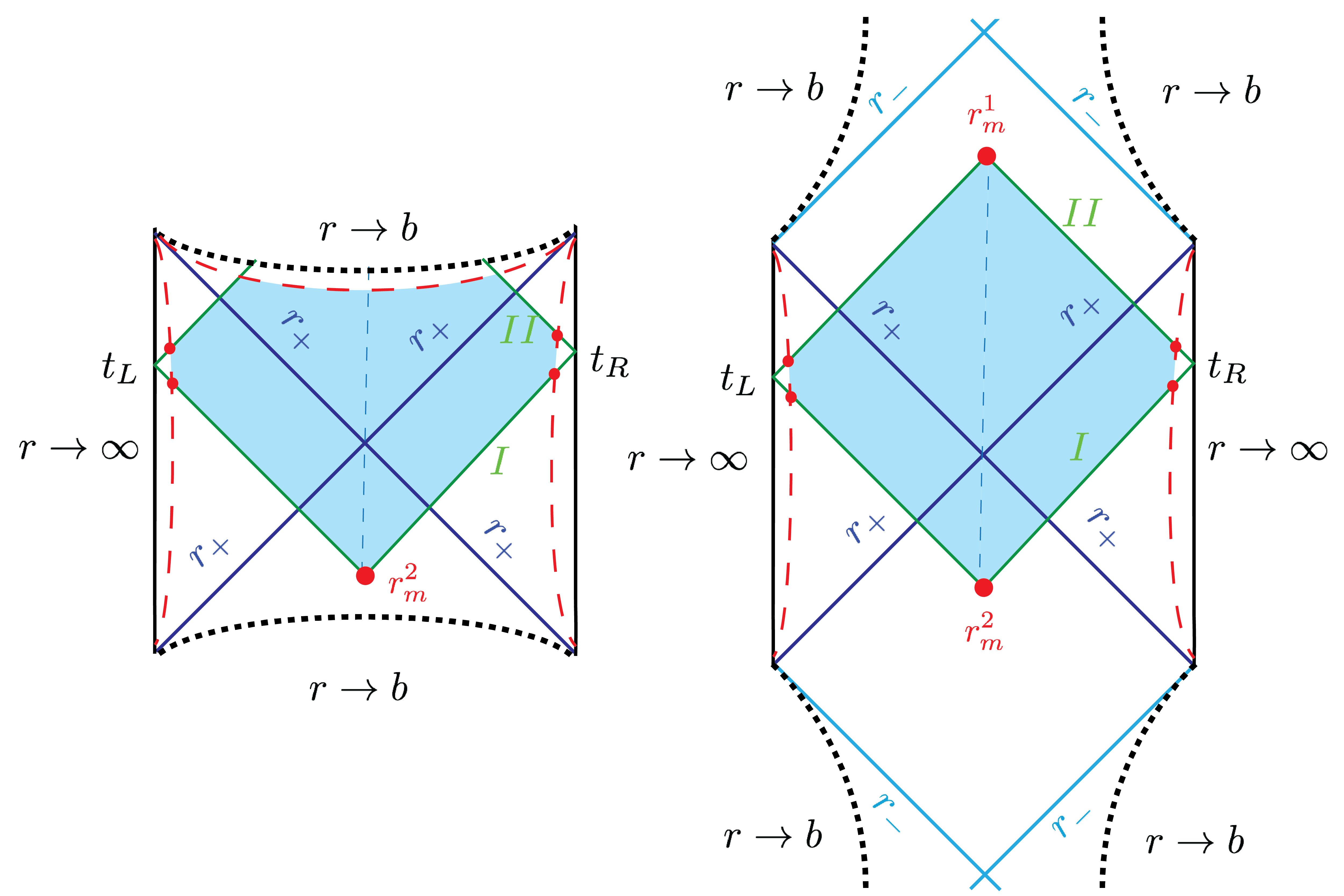}
\caption{Causal structure for the charged dilatonic black hole given by eq.~\eqref{metric_dilaton}. The left panel corresponds to $\alpha^2 \geq 1/{3}$, for which the causal structure is similar to that of the Schwarzschild-AdS black hole, with a spacelike singularity at $r = b$. The right panel corresponds to $0\le\alpha^2 < 1/{3}$, for which the causal structure is similar to that of the Reissner-Nordstrom-AdS black hole and the timelike singularity lies behind an inner Cauchy horizon (at $r=r_-$). }
\label{CausalDilaton}
\end{figure}

Implicitly, for the following, we will only consider nonextremal solutions, with $b$ positive and $c$ sufficiently large, \eg $c\gg b$. The causal structure for these solutions  is illustrated in figure \ref{CausalDilaton}. The geometry has a curvature singularity at $r=b$ where $U(r)$ vanishes with any finite $\alpha$. In general, there are horizons determined by $f(r_\pm)=0$. However, for $\alpha^2 \ge 1/{3}$, one generally finds a single (real) solution $r_+>b$ and the singularity is spacelike. Hence in examining the CA proposal, we will find the future null boundaries of the WDW patch meet the singularity (at late times), as illustrated in the left panel of figure \ref{CausalDilaton}.  Furthermore, for $0<\alpha^2<1/{3}$, there is an additional inner horizon at $r_{-}$ between the event horizon and the singularity at $r=b$, \ie $b<r_-<r_+$, as shown in the right panel of figure.\footnote{Of course, just as for the Reissner-Nordstrom-AdS solution, there is a threshold beyond which the charged dilatonic solution \reef{metric_dilaton} becomes a naked singularity, \eg if we begin with large $c$ but then reduce its value while holding $b$ fixed. For the theories with $0 < \alpha^2 < 1/3$, the threshold corresponds to the point where $r_{-}$ coincides with $r_{+}$, and hence the solution becomes an extremal black hole (matching the behaviour of the Reissner-Nordstrom-AdS black holes). However, the situation is different for $\alpha^2 \geq 1/3$ where the nonextremal black holes only have a single horizon. In this case, the threshold is reached when the event horizon meets the singularity, \ie $r_{+} \rightarrow b$, and hence the threshold solution contains a null singularity.}

Following \cite{Hendi:2010gq}, the mass of the black hole \reef{metric_dilaton} can be shown to be 
\begin{equation}
M=\frac{1}{2G_N}\left(c+\frac{1-\alpha^2}{1+\alpha^2}b\right)\,
\end{equation}
It is useful to use $f(r_+)=0$ to rewrite the parameter $c$ in terms of the position of the event horizon of the black hole,
\begin{equation}\label{c_parameter}
c= r_{+}+\frac{r_{+}^3}{L^2}\left(1-\frac{b}{r_{+}}\right)^{(3\alpha^2-1)/(1+\alpha^2)}\,.
\end{equation}
Then the temperature and entropy of the black hole can be expressed as
\begin{eqnarray}
T&=&\frac{1}{4\pi}\frac{\partial f}{\partial r}\bigg{|}_{r = r_+}=\frac{1}{4\pi r_{+}}\left(1-\frac{b}{r_{+}}\right)^{(1-\alpha^2)/(1+\alpha^2)}+\frac{3r_{+}(1+\alpha^2)-4b}{4\pi L^2(1+\alpha^2)}\left(1-\frac{b}{r_{+}}\right)^{(\alpha^2-1)/(1+\alpha^2)} \,,
\nonumber\\
S&=&\frac{\pi U^2(r_{+})}{G_N}=\frac{\pi}{G_N}r_{+}^2\left(1-\frac{b}{r_{+}}\right)^{2\alpha^2/(1+\alpha^2)}\,.
\label{DiTempEnt}
\end{eqnarray}
It would be interesting to study which black hole solutions are thermodynamically and dynamically stable (\eg in analogy to refs.~\cite{char1,char2}, however, we do not pursue this question here).

\subsection{Complexity Growth}
We will now study the time-dependence of the holographic complexity of the charged dilatonic black holes presented above using the CA proposal. Of course, in contrast to the previous discussion of the dyonic Reissner-Nordstrom-AdS black holes in section \ref{sec:RNBHS}, we only have the solutions carrying purely electric charges here. We will follow the discussion in \cite{Growth}, which is straightforward to adapt to these solutions. Further, we will only be considering the action \eqref{TOTDilaton} here and defer the discussion of additional boundary terms to the next sections. Since we are primarily interested in the late-time growth rate, for the theories with $\alpha^2\ge1/3$, we will assume that the WDW patch has already lifted off of the past singularity in the following calculations, as illustrated in left panel of figure \ref{CausalDilaton}.

\subsubsection*{Bulk Contribution}
Evaluating the bulk action \reef{pda} yields
\begin{align}
I_\mt{bulk}=&\frac{1}{2G_N}\int_\mt{WDW} dt dr\bigg[-\frac{r^2}{(1+\alpha^2)^2L^2}\bigg(8\alpha^2\left(1-\frac{b}{r}\right)^{\frac{3\alpha^2-1}{\alpha^2+1}}+(3-\alpha^2)\left(1-\frac{b}{r}\right)^{\frac{4\alpha^2}{\alpha^2+1}} \nonumber \\
&\qquad\qquad+\alpha^2(3\alpha^2-1)\left(1-\frac{b}{r}\right)^{2\frac{\alpha^2-1}{\alpha^2+1}}\bigg)+\frac{q_e^2}{r^2}\bigg]\,, \label{IntBulkDil}
\end{align}
The time derivative then becomes
\begin{equation}\label{dilaton_bulktime}
\frac{d I_\mt{bulk}}{dt}= \frac{1}{2G_N}\left[\frac{q_e^2}{r}+ \frac{r^2}{L^2 (1+\alpha^2)} \left( r(1+\alpha^2) - b\right) \left(1 - \frac{b}{r} \right)^{\frac{3 \alpha^2 -1}{1+ \alpha^2}} \right]_{r_m^2}^{r_m^1}\,.
\end{equation}
For black holes with just one horizon, \ie $\alpha^2 \geq 1/3$, $r_m^1$ corresponds to the position of the singularity, that is $r_m^1 = b$.  On the other hand, for $0<\alpha^2< 1/3$, the past meeting point approaches the Cauchy horizon at late times, \ie $r_m^1\to r_{-}$ -- see figure \ref{CausalDilaton}. Similarly, at late times, $r_m^2 \to r_+$ for all $\alpha$.

\subsubsection*{GHY contribution}

As noted above, for $0<\alpha^2<1/3$, the future tip of the WDW patch is the joint where the future null boundaries meet (with $r_-<r_m^1<r_+$). In contrast for $\alpha^2 \geq 1/3$, the WDW patch ends on the spacelike singularity at $r = b$ and so as usual, we introduce a regulator surface at $r=b+\epsilon_0$. We must evaluate the Gibbons-Hawking-York (GHY) term, given in eq.~\reef{surface}, on this surface and consider the limit $\epsilon_0\to0$. The trace of the extrinsic curvature of the regulator surface  is given by 
\begin{equation}
K=-\frac{1}{2\sqrt{-f(r)}}\left(\partial_r f(r)+2\frac{\partial_r(U(r)^2)}{U(r)^2}f(r)\right)\bigg|_{r=b+\epsilon_0}\,.
\end{equation}
However, notice that in integrating this term over the surface, the spherical measure is not $r^2$, \eg as in the Schwarzschild-AdS solution, but $U(r)^2$ instead. Hence the GHY contribution from the regulator surface becomes 
\begin{equation}
I_\mt{GHY}=-\frac{U(r)^2}{2G_N}\left(\partial_r f(r)+2\frac{\partial_r(U(r)^2)}{U(r)^2}f(r)\right) \left(\frac{t}{2}+r^{*}_{\infty}-r^*(r)\right)\bigg{|}_{r=b+\epsilon_0} \,.
\end{equation}
Now taking time derivative and the limit $\epsilon_0 \to 0$ yields
\begin{equation}\label{GHYDilaton}
\frac{d I_\mt{GHY}}{dt}=
\begin{cases}
\ \frac{3}{8G_N}\left(c-b-\frac{b^3}{L^2}\right)& \text{for}\,\, \alpha^2=\frac{1}{3}\\
\ \frac{(1+3\alpha^2)}{4G_N(1+\alpha^2)}\,(c-b)& \text{for}\,\, \alpha^2>\frac{1}{3}
\end{cases}
\end{equation}
Notice that subtleties in the $\epsilon_0 \rightarrow 0$ limit produce the extra term proportional to $b^3$ here when we have precisely $\alpha^2 = 1/3$.

\subsubsection*{Joint contributions}

If we focus on $\alpha^2 < 1/3$, the only joints which contribute to the time dependence are those at the future and past meeting points, \ie $r=r_m^1$ and $r_m^2$ -- see figure \ref{CausalDilaton}. The corresponding joint contributions are given by
\begin{equation}
I_{\text{joint}} (r_m^1) + I_{\text{joint}} (r_m^2)  = -\frac{U^2(r_m^1)}{2G_N}\log{\frac{|f(r_m^1)|}{\xi^2}} -\frac{U^2(r_m^2)}{2G_N}\log{\frac{|f(r_m^2)|}{\xi^2}} \, .
\end{equation}
The time derivative then yields
\begin{equation}\label{JointFun}
\frac{d}{d t} \left( I_{\text{joint}} (r_m^1) + I_{\text{joint}} (r_m^2)  \right) = \left[\frac{U^2(r)}{4G_N}\left(\partial_rf(r)+\frac{\partial_r(U^2(r))}{U^2(r)}f(r)\log\frac{|f(r)|}{\xi^2}\right)\right]_{r_m^1}^{r_m^2}
\end{equation} 
As discussed above for $\alpha^2 \geq 1/3$, the future boundary of the WDW patch is the regulator surface just above the spacelike singularity. While there are joints where the future null boundaries meet this surface, their size is proportional to $U^2(r=b+\epsilon_0)$ which vanishes in the limit $\epsilon_0\to0$. Hence the corresponding joint contributions vanish. Therefore in this case, the contribution to the time derivative comes from the past meeting point and it is precisely given by the expression above evaluated at  $r=r_m^2$.

\subsubsection*{Counterterm contribution}
To evaluate the surface counterterm \eqref{Counterterm}, we begin by choosing the affine parameter along the null boundaries as
\begin{equation}
\lambda=\frac{r}{\xi}\,,
\end{equation}
which then yields 
\begin{equation}
\Theta=\frac{\xi \partial_r (U(r)^2)}{U(r)^2}\,.
\end{equation}
The sum of the counterterm contributions on the four null boundaries then reads 
\begin{equation}
I_\mt{ct}=\frac{1}{G_N}\int_{r_m^1}^{r_{max}} dr \,\partial_r (U(r)^2) \log{\frac{\xi \ctL \partial_r (U(r)^2)}{U(r)^2}}+\frac{1}{G_N}\int_{r_m^2}^{r_{max}} dr\, \partial_r (U(r)^2) \log{\frac{\xi \ctL \partial_r (U(r)^2)}{U(r)^2}}\,.
\end{equation}
This integration is nontrivial for general values of $\alpha$. However, the time dependence has a simple form, 
\begin{equation}\label{CTfun}
\frac{d I_{ct}}{dt}=\frac{1}{2 G_N}\left[\partial_r(U(r)^2)f(r)\log{\frac{\xi \ctL \partial_r(U(r)^2)}{U(r)^2}}\right]_{r_m^1}^{r_m^2}\,.
\end{equation}
Implicitly, the contribution evaluated at $r_m^1$ would absent at late times if we consider the solutions for $\alpha^2 \geq 1/3$. As expected, when eqs.~\eqref{JointFun} and \eqref{CTfun} are added together, the combined contribution to the time derivative is independent of $\xi$.

\subsubsection*{Total growth rate}
Now we combine all of the contributions in eqs.~\eqref{dilaton_bulktime}, \eqref{GHYDilaton}, \eqref{JointFun} and \eqref{CTfun} and consider the late-time limit, to find
\begin{equation}\label{GrowthDilaton}
\lim_{t \to \infty}\frac{d \mathcal{C}_{A}}{dt}=
\begin{cases}
\frac{q_e^2}{\pi G_N}\left[\frac{1}{r_{-}}-\frac{1}{r_{+}}\right] & \mbox{for} \,\,  \alpha^2<\frac{1}{3}\\
\frac{1}{\pi}\left[2M-\frac{q_e^2}{ G_N r_{+}}-\frac{3b}{4G_N}-\frac{b^3}{4G_NL^2}\right]& \mbox{for}\,\, \alpha^2=\frac{1}{3}\\
\frac{1}{\pi}\left[2M-\frac{q_e^2}{ G_N r_{+}}-\frac{b}{(1+\alpha^2)G_N}\right]& \mbox{for}\,\, \alpha^2>\frac{1}{3}
\end{cases}
\end{equation}
Of course, the late-time growth rate depends  on the causal structure of the black hole -- see figure \ref{CausalDilaton}. In particular, we note that in the theories with $0 \leq \alpha^2<1/3$ for which the causal structure matches that of the Reissner-Nordstrom-AdS black holes, the form of the late-time rate above has precisely the same form as in eq.~\reef{square} for the latter solutions. In fact, the result in eq.~\eqref{GrowthDilaton} reduces to precisely the growth rate of the (electrically charged) Reissner-Nordstrom-AdS black holes when $\alpha \to 0$. We also note that in the limit $\alpha \to \infty$, we recover the late-time growth rate of the Schwarzschild-AdS solution, \ie $d \mathcal{C}_{A}/dt=2M/\pi$. Further, we observe that
using \eqref{c_parameter}, this rate will vanish as we approach extremality, \ie as $r_{+}\to r_{-}$ for $0 \leq \alpha^2<1/3$, which again parallels the behaviour of the (electrically charged) Reissner-Nordstrom-AdS black holes \cite{Growth,Brown:2018kvn}.
We also note that for the $\alpha^2 \geq 1/3$ solutions, the rate vanishes in the limit $r_{+} \to b$, where the black holes become null singularities. 

\begin{figure}
\centering
\includegraphics[scale=1]{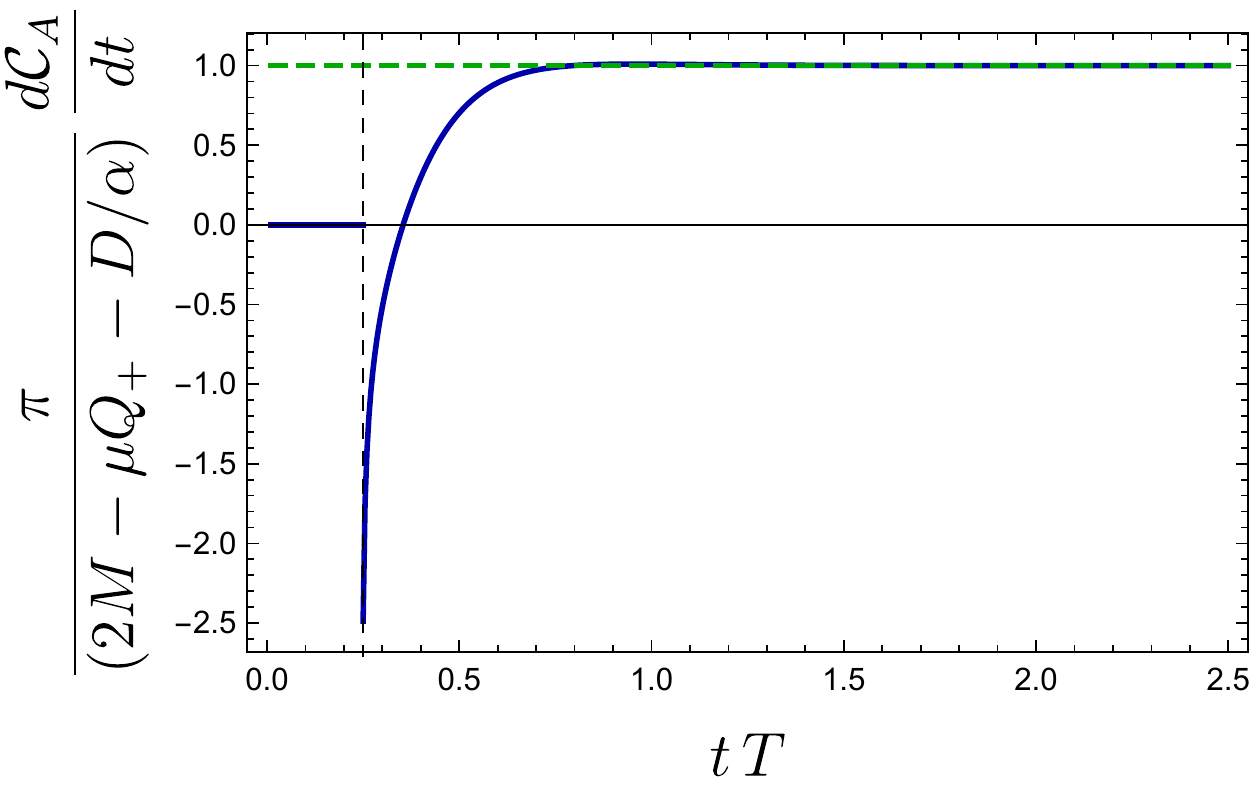}
\caption{The time dependence of complexity for an electrically charged Maxwell-Dilaton black hole (without the addition of the Maxwell boundary term). We evaluate for concreteness $\alpha=1/\sqrt{2}$, which corresponds to a black hole with a causal structure that resembles that of an Schwarzschild-AdS black hole. The parameters are chosen to be $\ell_\mt{ct} = L$, $L=0.9$, $b=0.75$ and $c=2.5$. In analogy to the Schwarzschild-AdS black hole, the complexity does not change until a certain critical time, where the WDW patch leaves the past singularity. Then, the complexity approaches the late time limit from above, with a transient dependence on $\ell_\mt{ct}$, at times of the order of the inverse temperature.  }
\label{MaxDil1sqrt2}
\end{figure}

As an example, we show in figure \ref{MaxDil1sqrt2} the full time evolution of complexity for $\alpha^2 =1/2$, for which the causal structure resembles that of an Schwarzschild-AdS black hole (left panel in figure \ref{CausalDilaton}). The behaviour is very similar to that of the latter neutral black holes, as shown in the detailed analysis of \cite{Growth}. Up to a certain critical time, the WDW patch ends on both the past and future singularities, and during this time, the complexity remains constant. After this critical time, the past null boundaries meet at $r=r_m^2$, as discussed above, and at late times, this joint approaches the event horizon. In this period of time, rate of change of the complexity exhibits a transient behaviour (which depends on the counterterm scale $\ell_\mt{ct}$) and then by a time of the order of the inverse temperature, it has overshot the late-time limit which it subsequently approaches from above.

\subsection{Boundary Terms} \label{blt2}
Next we examine how the growth rate of the holographic complexity \reef{defineCA} for the charged dilatonic black holes \reef{metric_dilaton} is effected by the addition of two boundary terms, involving the Maxwell and dilaton fields.

\subsubsection*{Maxwell Boundary Term}
We begin with the Maxwell boundary term for the new theory \reef{pda}, 
\begin{equation}\label{mbadila}
I_{\mu \mt{Q}}=\frac{\gaa}{g^2}\int_{\partial\mathcal{M}}d\Sigma_\mu\, F^{\mu \nu}\,A_\nu\, e^{-2\alpha\phi}\,,
\end{equation}
where $\gaa$ is a free parameter.
Following the same reasoning as in section \ref{sec:MBA}, this boundary term changes the boundary condition imposed on the Maxwell field in the variational principle. However, we should add that implicitly we would also be assuming a Dirichlet boundary condition for the dilaton, \ie $\delta \phi|_{\partial{\cal M}}=0$.
Further in analogy with \eqref{MaxBdryActOnS}, if the Maxwell field satisfies the equation of motion $\nabla_\mu(e^{-2 \alpha \phi} F^{\mu\nu})=0$, this boundary term is equivalent to 
\begin{equation}
I_{\mu\mt{Q}}\big{|}_{\mt{on shell}}=\frac{\gaa}{2g^2}\int_\mathcal{M}d^4x \sqrt{-g}\, e^{-2 \alpha \phi}\,F^{\mu \nu}F_{\mu \nu}\,.
\end{equation}
Hence, it is straightforward to evaluate the effect of this boundary term \eqref{mbadila} on the time dependence of the WDW action and one finds
\begin{equation}\label{late}
\frac{dI_{\mu \mt{Q}}}{dt}= -\frac{\gaa\,q_e^2}{G_N}\left[ \frac{1}{r_m^2} -\frac{1}{r_m^1}\right]
\end{equation}
Of course, all of the contributions calculated previously are unchanged. Hence adding in the above expression, the late time limits in eq.~\reef{GrowthDilaton} are now replaced by
\begin{equation}\label{GrowthDilaFix}
\lim_{t \to \infty}\frac{d \mathcal{C}_{\mathcal{A}}}{dt}=
\begin{cases}
\frac{(1-\gaa)q_e^2}{\pi G_N}\left[\frac{1}{r_{-}}-\frac{1}{r_{+}}\right] & \mbox{for} \,\, \alpha^2<\frac{1}{3}\\
\frac{1}{\pi}\left[2M-\frac{(1-\gaa)q_e^2}{G_N r_{+}}-\frac{3(b+\gaa c)}{4G_N}-\frac{b^3}{4G_NL^2}\right]& \mbox{for}\,\, \alpha^2=\frac{1}{3}\\
\frac{1}{\pi}\left[2M-\frac{(1-\gaa)q_e^2}{G_N r_{+}}-\frac{b+\gaa c}{(1+\alpha^2)G_N}\right]& \mbox{for}\,\, \alpha^2>\frac{1}{3}
\end{cases}
\end{equation}
Notice that, if we fix $\gaa$, the limits $\alpha \to \infty$ and $\alpha \to 0$ discussed below eq.~\eqref{GrowthDilaton} are unchanged, \ie they yield the late-time growth rates of the Schwarzschild-AdS and electrically charged Reissner-Nordstrom-AdS black holes, respectively. Further, as before, the above rate will vanish as we approach extremality for $0 \leq \alpha^2<1/3$, and as we approach the limit of a null singularity for $\alpha^2 \geq 1/3$. 

One interesting choice to consider for the boundary coefficient is $\gaa=1$, for which eq.~\eqref{GrowthDilaFix} becomes
\begin{equation}\label{DilaNewmann}
\lim_{t \to \infty}\frac{d \mathcal{C}_{\mathcal{A}}}{dt}\bigg{|}_{\gaa=1}=
\begin{cases}
0  & \mbox{for} \,\, \alpha^2<\frac{1}{3}\\
\frac{1}{2\pi}\left[M-\frac{3b}{4G_N}-\frac{b^3}{2G_NL^2}\right]& \mbox{for}\,\, \alpha^2=\frac{1}{3}\\
\frac{1}{\pi}\left[\frac{2\alpha^2}{1+\alpha^2}\left(M-\frac{b}{(1+\alpha^2)G_N}\right)\right]& \mbox{for}\,\, \alpha^2>\frac{1}{3}
\end{cases}
\end{equation}
That is, for $0\leq \alpha^2<\frac{1}{3}$ in which case the causal structure matches that of the Reissner-Nordstrom-AdS black holes, the (electrically) charged dilatonic black holes fail to complexify at late times. This precisely matches the behaviour found in section \ref{sec:MBA}. On the other hand, for $\alpha^2>\frac{1}{3}$ in which case the causal structure is similar to the Schwarzschild-AdS black holes, the late-time growth rate remains nonvanishing. However, we observe that in the uncharged limit (\ie $b \to 0$), eq.~\eqref{DilaNewmann} does not yield the expected growth rate of $2M/\pi$ -- see section \ref{sec:Discussion} for further discussion. 

\subsection*{Dilaton Boundary term} 
Next we consider the following boundary term for the dilaton 
\begin{equation}\label{dilaton_change}
I_{\phi}=\frac{\gamma_\phi}{4\pi G_N}\int_{\partial \mathcal{M}} d\Sigma_\mu\,\phi\, \partial^\mu \phi \,.
\end{equation}
As for the Maxwell boundary term \reef{mbadila} (or eq.~\reef{mba2} in the previous section), this term modifies the character of the boundary condition which must be imposed on the dilaton in the variational principle. For example, while $\gamma_\phi=0$ corresponds to a Dirichlet boundary condition (\ie $\delta \phi|_{\partial{\cal M}}=0$), setting $\gamma_\phi=1$ yields a Neumann boundary condition (\ie $n^\mu\partial_\mu\delta \phi|_{\partial{\cal M}}=0$). More general choices of this parameter lead to mixed boundary conditions. Further, if both $\gamma_\phi$ and $\gaa$ are nonvanishing, the dilaton will have a more complicated boundary condition involving terms proportional to the integrand in eq.~\reef{mbadila}.

Let us first consider black holes for $0<\alpha^2<1/3$, in which the causal structure resembles the Reissner-Nordstrom-AdS black hole as shown in figure \ref{CausalDilaton}. In this case, the boundary term lives only on the null boundaries of the WDW patch but in this case, the derivative appearing in eq.~\reef{dilaton_change} is actually tangent to the boundary. Therefore the boundary term reduces to an integral over the joints where the null boundaries intersect, namely
\begin{equation}
I_{\phi}=\frac{\gamma_\phi}{8\pi G_N}\int_{\Sigma '} d^2x \sqrt{\sigma} \phi^2\,,
\end{equation}
where each joint term carries a sign according to the conventions of \cite{NullBound}. However, one finds in this case
\begin{equation}
\lim_{t \to \infty} \frac{d I_{\phi }}{dt}=0\,.
\end{equation}
Hence, adding the dilaton boundary term \eqref{dilaton_change} does not change the complexity growth rate at late times for these black holes. Nevertheless, the transient behaviour of the holographic complexity at early times will be modified by this term, but we will not explore this here.

Next, we turn to the case $\alpha^2\ge 1/3$, in which the causal structure resembles the Schwarzschild-AdS black hole. In this case, the contribution from the null boundaries of the WDW patch still reduce to contributions on various joints, which again do not contribute to the late-time growth rate. However, there is an additional contribution coming from the (spacelike) regulator surface at the future singularity -- see figure \ref{CausalDilaton}. Evaluating eq.~\eqref{dilaton_change} on this boundary and considering the late time limit, we find
\begin{equation}
\lim_{t \to \infty} \frac{dI_{\phi}}{dt}=\lim_{\epsilon_0 \to 0} \gamma_\phi\,\frac{U(r)^2f(r)\,\phi\,\partial_r \phi}{2G_N}\bigg{|}_{r = b+\epsilon_0}\,.
\end{equation}
Unfortunately, for $\gamma_\phi \neq 0$, this expression is divergent. Therefore adding the dilaton boundary term \eqref{dilaton_change} spoils the good behaviour of the regularization procedure at the singularity. Therefore, we do not consider these boundary terms further here.\\

Hence, our general results for the late-time growth rate of the holographic complexity including the Maxwell boundary term \eqref{mbadila} are summarized eq.~\reef{GrowthDilaFix} for the electrically charged black holes. Of course, these results match the growth rates without the Maxwell boundary term in eq.~\reef{GrowthDilaton} when we set $\gamma_\alpha=0$. However, when the Maxwell boundary term \eqref{mbadila} was included, we also showed in eq.~\reef{DilaNewmann} that choosing $\gamma_\alpha=1$ sets the late-time growth rate to zero for the cases where the causal structure was like that of the Reissner-Nordstrom-AdS black holes, \ie for $\alpha^2<1/3$. No such choice was possible when the causal structure had the form of the Schwarzschild-AdS black holes, \ie for $\alpha^2\ge 1/3$. The former behaviour was analogous to that found in the Einstein-Maxwell theory in section \ref{sec:RNBHS} and therefore it appears that the causal structure of the black hole was one of the essential features producing the unusual behaviour found there.
However, we note that our analysis here focused only on electrically charged black holes and we did not consider dyonic or magnetically charged black holes. Unfortunately the latter solutions are not yet known for the Einstein-Maxwell-Dilaton theory \reef{pda}. We return to this point in section \ref{sec:Discussion}.

\section{Black Holes in Two Dimensions}\label{sec:JT}

In this section, we will focus on studying dilaton gravity models in two bulk spacetime dimensions. Our main motivation is evaluating the growth of holographic complexity for the Jackiw-Teitelboim (JT) model \cite{Jackiw:1984je,Teitelboim:1983ux, AP}, which has a simple action linear in the dynamical dilaton field. This theory has received great deal of attention recently as the gravitational dual of the Sachdev-Ye-Kitaev (SYK) model in the low energy limit, where the system acquires an emergent reparametrization invariance \cite{PhysRevLett.70.3339, Kitaev, Kitaev2,
Plochinski:2016,Maldacena:2016hyu,Maldacena:2016upp}. One perspective of JT gravity is that it describes physics (of the spherically symmetric sector) in the near-horizon region of near-extremal charged black holes in higher dimensions, \eg \cite{Sen:2007qy,Kunduri:2007vf, Kunduri:2013ana, Castro:2008ms, Nayak:2018qej, Almheiri:2016fws, Sarosi:2017ykf}. More specifically, we focus on deriving the action for JT gravity by reducing the action \eqref{TOTaction} to two dimensions while assuming the background is spherically symmetric and magnetically charged in four dimensions, \ie the four-dimensional gauge field has the form \eqref{FieldDyon} with $q_e=0$. In addition in this section, we will  analyze an analogous two-dimensional theory that can describe the near-horizon physics of four-dimensional black holes carrying a purely electric charge, \ie eq.~\eqref{FieldDyon} with $q_m=0$. The two-dimensional Maxwell field is an essential ingredient for this JT-like theory and so it has a form reminiscent of the Brown-Teitelboim model \cite{BT1,BT2}, where the effective cosmological constant is dynamically controlled by the energy density of an antisymmetric $d$-form field strength in $d$ dimensions. Further, our analysis of holographic complexity in the previous sections has shown the important role of the Maxwell boundary term \reef{mba0}. Hence while we begin by examining the dimensional reduction without this term, \ie by reducing $I_\mt{0}$ in eq.~\reef{Inot}, we also consider the dimensional reduction of this boundary term and its contribution to  the holographic complexity for both the JT and JT-like models.
As might be expected, we will find the holographic complexity for both models behaves in the same way as for the corresponding four-dimensional black holes discussed section \ref{sec:RNBHS}. We will discuss these theories and the holographic complexity in more detail in an upcoming work \cite{HolographicJT}.

\subsection{Jackiw-Teitelboim Model} \label{wonderland}

We begin with the dimensional reduction of the action \eqref{TOTaction} but without the addition of the Maxwell boundary term, \ie setting $\gamma=0$ \cite{Trivedi1992,Lechtenfeld,StTr,FIS}. We decompose the four-dimensional metric as 
\begin{equation} \label{metric_ansatz}
ds^2=g_{ab}(x)\,dx^a\, dx^b + \Psi^2 \left(d\theta^2+\sin^2\theta d\phi^2 \right)\, .
\end{equation}
If we assume that the  Maxwell field in four dimensions corresponds to a pure magnetic charge, we can use this metric ansatz to solve for $F$, and the result is precisely that given by eq.~\eqref{FieldDyon} with $q_e = 0$.
Substituting eq.~\eqref{metric_ansatz} and this magnetic field into the bulk action \eqref{bulk}, we integrate out the spherical directions to produce the following two-dimensional action
\begin{equation}\label{action_preJT}
\begin{split}
I_\mt{mag}^\mt{2D} &=\frac{1}{4 G_N} \int_\mathcal{M} d^2x \sqrt{-g}\left(\Psi^2 \mathcal{R} + 2 \left(\nabla \Psi \right)^2 - U(\Psi)\right)\\
& \qquad\qquad + \frac{1}{2 G_N} \int_{\partial \mathcal{M}} d x  \, \sqrt{|\gamma|} \, n^{\mu} \nabla_{\mu} \Psi^2  \, ,
\end{split}
\end{equation}
with the potential given by
\begin{equation}\label{potentialJT}
U(\Psi)=-2-6\frac{\Psi^2}{L^2}+2\frac{q_m^2}{\Psi^2}\,.
\end{equation}
The boundary term in the second line of eq.~\reef{action_preJT} results from integrating by parts in the dimensional reduction. We emphasize that it arises from the bulk terms \reef{bulk} in the four-dimensional action and is unrelated to the surface terms \reef{surface} or the null counterterm \reef{Counterterm}, whose dimensional reduction we will explicitly examine below. The action  \eqref{action_preJT} illustrates the fact that restricted to spherically symmetric solutions, our theory can be recast as a two-dimensional gravity model with a dilaton field. However, no approximations have been made at this point, and so the full four-dimensional solution \reef{RNBas} can be recovered from eq.~\eqref{action_preJT}.

Next, we are interested in describing the near-horizon region of the near-extremal black holes. Recall that in extremal limit, the charged black holes develop an infinitely long throat of a fixed radius $r_h$ \cite{Nayak:2018qej}. That is, the near-horizon region of the extremal solutions is described by a constant dilation profile $\Psi^2=r_h^2$. For latter purposes, we define the extremal horizon area as 
\begin{equation}\label{constant_PHI0}
\Phi_0\equiv 4 \pi r_h^2 \, .
\end{equation}
For the extremal solutions, we have $f_{\RNo}(r_h)=0=f'_{\RNo}(r_h)$ which allows us to express the extremal charge in terms of the horizon radius,
\begin{equation}\label{extremal_charge}
\qtx^2=r_h^2\left(1+3\frac{r_h^2}{L^2}\right)\,.
\end{equation}
Further, in the extremal throat, the two-dimensional geometry described by $g_{ab}$ has a constant negative curvature, which is related to the higher dimensional parameters by
\begin{equation}\label{Lambda2Eq}
\Lambda_2 =  - \frac{1}{L_2^2}  =-\left(\frac{1}{r_h^2}+\frac{6}{L^2}\right)\,.
\end{equation}

\begin{figure}
\centering
\includegraphics[scale=0.75]{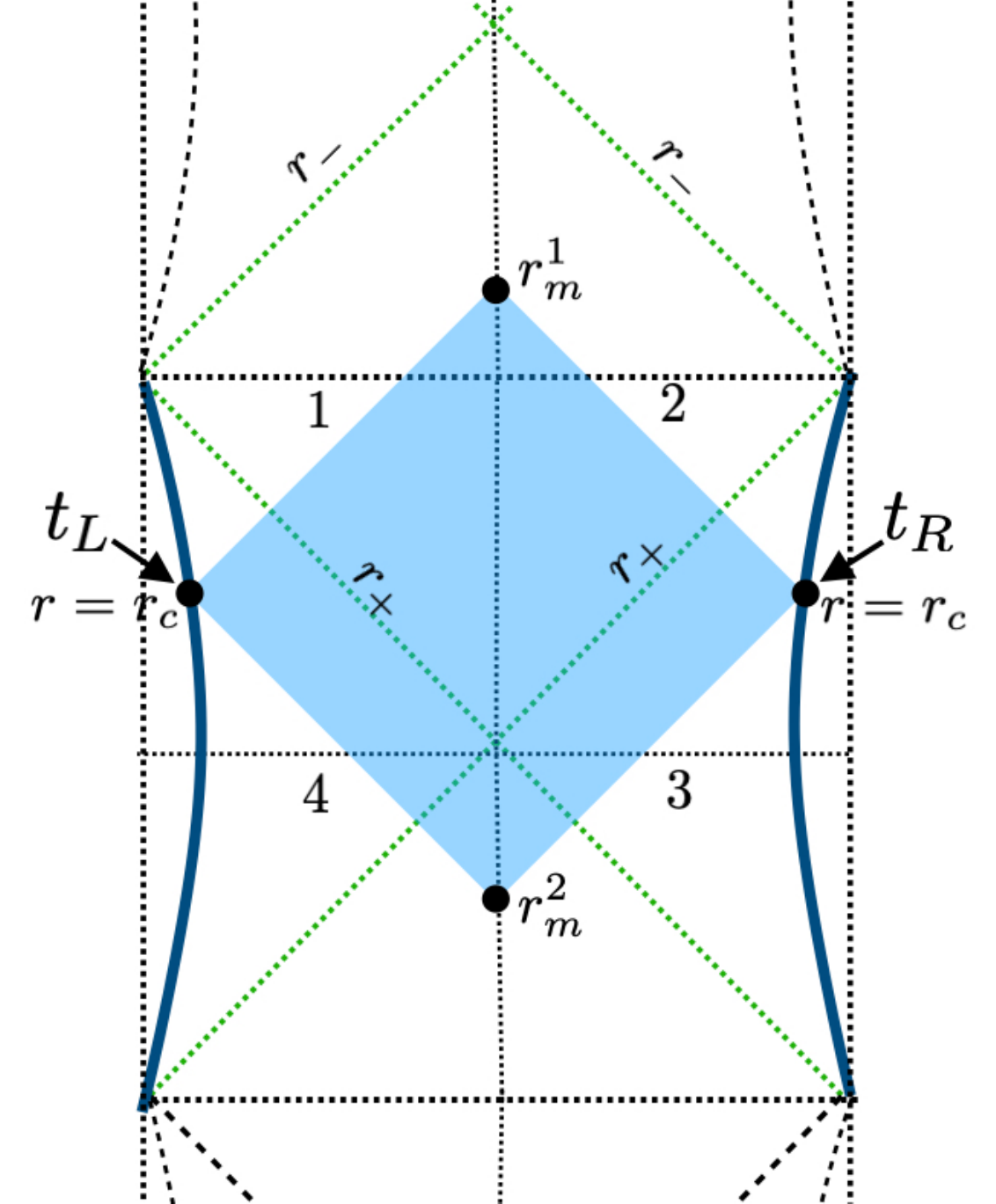}
\caption{AdS$_2$ solution of the JT model and the WDW patch. Physical boundary is depicted with a blue curve. The outer and inner horizons appear at $r=r^\mt{JT}_{\pm}=\pm\mu$.}
\label{WDWpatch}
\end{figure}

Now, in considering small deviations from the extremal throat, we expand the dilaton around the extremal value in eq.~\reef{constant_PHI0}. That is, we write 
\begin{equation}\label{small}
\Psi^2= \frac{1}{4 \pi} \, \left( \Phi_0+\Phi \right) \,,
\end{equation}
with the understanding that $\Phi/\Phi_0\ll1$.
In particular, applying this expansion (to linear order in $\Phi$) to the action \eqref{action_preJT} yields the Jackiw-Teitelboim action, 
\begin{equation}\label{JT_Action}
I^{\mt{JT}}_{\mt{bulk}} = \frac{\Phi_0}{16 \pi G_N}\int_{\mathcal{M}} d^2x \sqrt{-g}\, \mathcal{R} + \frac{1}{16\pi G_N}\int_{\mathcal{M}} d^2x \sqrt{-g}\,\Phi\left(\mathcal{R}- 2 \Lambda_2 \right) \, .
\end{equation}
The solutions derived from this action can be written as
\be
\Phi=\Phi_b\f{r}{r_c}\,,
\qquad ds^2=-f(r)dt^2+\f{dr^2}{f(r)}\, \quad \text{with} \ \  f(r)\equiv\f{r^2-\mu^2}{L_2^2} \, .\label{AdScoord}
\ee
In the dilaton solution, we have introduced the cut-off radius $r_c$. As depicted in Figure \ref{WDWpatch}, this time-like surface $r=r_c$ determines the position of the physical boundary of our system. The dynamics of the boundary position reproduces the IR physics of the SYK model, as has extensively been studied in recent years \cite{Kitaev, Kitaev2,
Plochinski:2016,Maldacena:2016hyu,Maldacena:2016upp, Jensen:2016pah,EMV}.  The boundary value of the dilaton is denoted $\Phi_b$ and the linear approximation remains valid as long as $\Phi_b/\Phi_0\ll 1$.
The metric has an outer and inner horizon at $r_\pm^\mt{JT}=\pm{\mu}$. The black hole is characterized by the following parameters 
\begin{equation}
M_\mt{JT} = \frac{\Phi_b \,\mu^2 }{16\pi G_N L_2^2 \,r_c}\, , \qquad\  S_\mt{JT}=\frac{\Phi_0+\Phi(r_+^\mt{JT}=\mu)}{4 G_N}  \, ,\ \qquad T_\mt{JT} = \frac{{\mu}}{2\pi L_2^2}\, . \label{MSTJT}
\end{equation}
The mass $M_\mt{JT}$ and temperature $T_\mt{JT}$ are taken as energies conjugate to the coordinate time $t$ (which will be taken as the time in the boundary theory). Of course, one can treat the JT model as an independent theory, or one can match the JT solutions \reef{AdScoord} with a description of the near-extremal throats  of the Reissner-Nordstrom-AdS black holes \reef{RNBas} (within the linear approximation applied above).

\subsubsection*{Complexity Growth}
Next, we consider the growth of holographic complexity for the JT model using the CA proposal. As depicted in Figure \ref{WDWpatch}, we consider  the WDW patch anchored on the physical boundary $r=r_c$. As in section \ref{sec:RNBHS} following \cite{Growth}, we anchor this region at the boundary times $t_L=t_R=t/2$. Further, as illustrated in the figure, we denote the meeting points of the future and past null boundaries as $r=r^1_m$ and $r=r^2_m$, respectively. 

\begin{figure}
\centering
\includegraphics[height=6cm]{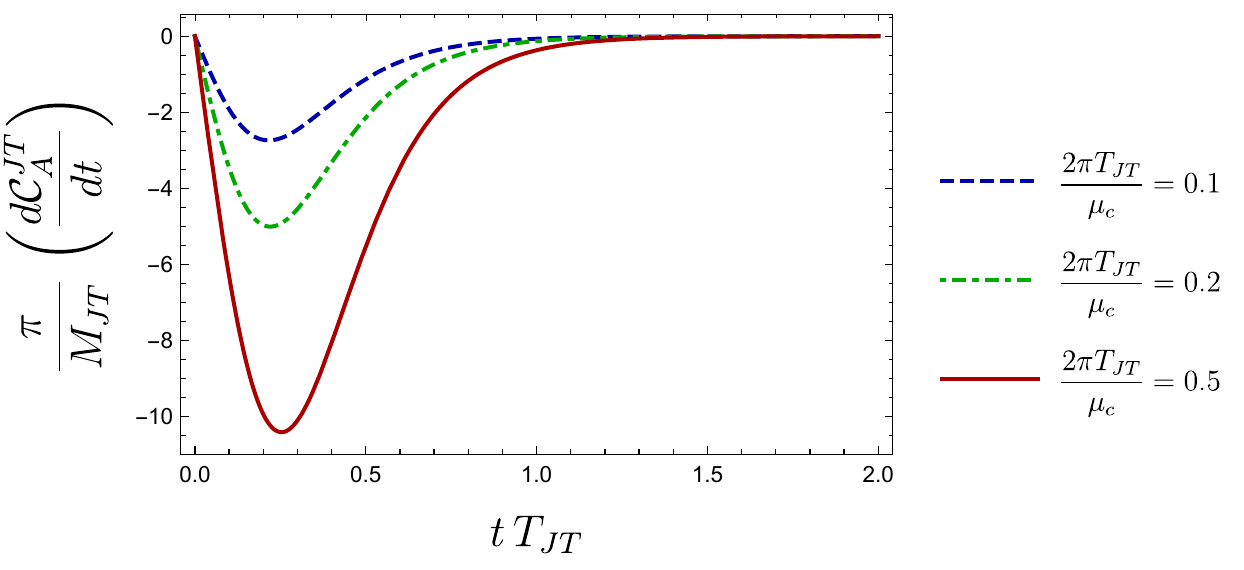}
\caption{Complexity growth in the JT model from eq.~\reef{Complexity_JT_dimless}. Different colours correspond to different temperatures. We set $\frac{\Phi_b}{\Phi_0} = 10^{-3}$ and $\ell_{ct}=L_2$. As we can see, the curves are qualitatively similar to the purely magnetic curve in \ref{Fig:DyonicEter}, with a negative transient behaviour, and the vanishing late time rate of change.}
\label{JTCA}
\end{figure}

First, we evaluate eq.~\eqref{JT_Action} on the WDW patch, which yields
\bal
I^\mt{ JT}_\mt{ bulk}=\left[\f{\Phi_0}{8\pi G_N}\log |f(r)|\right]^{r^1_m}_{r_c}+\left[\f{\Phi_0}{8\pi G_N}\log |f(r)|\right]^{r^2_m}_{r_c}\, . \label{Bulk_JT}
\eal
However, recall that the reduced action \eqref{action_preJT} included a surface term which was not incorporated in the JT action \eqref{JT_Action}.\footnote{One might also wish to consider JT gravity \eqref{JT_Action} in its own right,  without any reference to higher dimensions. In this case, we would not include the total derivative contribution \reef{hop1} as part of the WDW action. However, dropping this contribution would not change the vanishing growth rate \reef{JT_NoGrowth} at late times, but the transient behaviour for $t T_\mt{JT}\lesssim 1$ would be slightly modified, \eg in figure \ref{JTCA}.} Substituting eq.~\reef{small} in this surface term, we find a term that is linear in $\Phi$ and when it is evaluated on the null boundaries of the WDW patch, this surface term yields
\begin{equation}
I^\mt{JT}_{\mt{totder}} = \frac{1}{8 \pi G_N}\int_{\mathcal{B}^{'}} \, d \lambda\, k^{\mu} \, \nabla_{\mu} \Phi  = \frac{\Phi}{4 \pi G_N}  \bigg{|}^{r_m^1}_{r_c} +\frac{\Phi}{4 \pi G_N}  \bigg{|}^{r_m^2}_{r_c} \,  .
\label{hop1}
\end{equation}
Here, we are assuming that the null boundaries are affinely parametrized, with the null normal normalized as $k^\mu\partial_\mu=\partial_\lambda$. 

The remaining boundary terms introduced in eqs.~\reef{surface} and \reef{Counterterm} have a simple dimensional reduction. Of course, with affine parametrization along the null boundaries of the WDW patch, we have $\kappa=0$ and we can ignore the corresponding surface term in eq.~\eqref{surface}. This leaves only the null joint terms (proportional to $a$) and the null surface counterterm. Dimensionally reducing these two terms and substituting eq.~\reef{small} then yields the relevant boundary terms for the JT model,
\begin{align}
& I^{\mt{JT}}_{\mt{joint}} = \frac{1}{8 \pi G_N} \, \sum_i \left( \Phi_0 + \Phi \right) a \big|_{\Sigma'_i} \, ,  \nonumber \\
&I^{\mt{JT}}_{\mt{ct}} = \frac{1}{8 \pi G_N} \, \int_{\mathcal{B}^{'}} d \lambda \,  \left( \Phi_0 + \Phi \right)  \, \Theta_{2d} \log \left( \ell_{\mt{ct}} \Theta_{2d} \right) , \label{JTjntct}
\end{align}
where the two-dimensional `scalar expansion' reads $\Theta_{2d}  = \partial_{\lambda} \log \left( \Phi_0 + \Phi \right)$. Combining these surface terms with eq.~\reef{hop1} yields the total of the boundary contribution for the WDW patch illustrated in figure \ref{WDWpatch},
\bal
I^\mt{JT}_{\mt{totder}} + I_{\mt{joint}}^{\mt{JT}} + I_{\mt{ct}}^{\mt{JT}} 
=&\frac{1}{8\pi G_N}\left[2\Phi-(\Phi_0+\Phi)\,\log\!\left(\f{\ell^2_{ct}\Phi_{b}^{2}|f(r)|}{r_c^2\,\Phi_0^2}\right)\right]^{r^1_m}_{r_c}\no
 &+\frac{1}{8\pi G_N}\left[2\Phi-(\Phi_0+\Phi)\,\log\!\left(\f{\ell^2_{ct}\Phi_{b}^{2}|f(r)|}{r_c^2\,\Phi_0^2}\right)\right]^{r^2_m}_{r_c}\, .\label{Joint_JT}
\eal
Adding this expression to eq.~\reef{Bulk_JT} in eq.~\reef{defineCA} yields the holographic complexity for the CA proposal.

As the higher dimensional calculations in section \ref{subsec:CompDyon}, the complexity growth rate is determined by the dynamics of the meeting points, $r_m^1$ and $r_m^2$, of the future and past null boundaries of the WDW patch. In analogy to eq.~\eqref{MeetPntRN}, we find $dr_m^1/dt=f(r_m^1)/2$ and
$dr_m^2/dt=-f(r_m^2)/2$ where $f(r)$ is defined in eq.~\reef{AdScoord}. The growth rate of the holographic complexity is then given by
\bal \label{Complexity_JT}
\f{d{\cal C}^{\mt{JT}}_{A}}{dt}&= -\f{\Phi_b }{16\pi^2 \, r_c \, G_N}\left[f(r)\log\left(\f{\ell_{ct}^2\Phi_{b}^{2}|f(r)|}{ r_c^2 \, \Phi_0^2}\right)\right]^{r_m^1}_{r_m^2}\, .
\eal
At late times, $r^1_m$ and $r^2_m$  approach the inner and outer horizons, respectively, \ie $r^1_m\r r^{\mt{JT}}_-=-\mu$ and $r^2_m\r r^{\mt{JT}}_+=\mu$. Hence the prefactor of $f(r)$ in eq.~\reef{Complexity_JT} is vanishing in both contributions and thus we have
\begin{equation}\label{JT_NoGrowth}
 \lim_{t \to \infty}\frac{d \mathcal{C}^{\rm JT}_{A}}{d t}   = 0   \, .
\end{equation}

It is interesting to recast eq.~\eqref{Complexity_JT} into an expression involving the boundary or physical parameters of the JT model, \eg the mass, temperature and entropy in eq.~\eqref{MSTJT}. First, we define dimensionless coordinates for the meeting points.
\beqa
x_{m}^{1} &\equiv& \frac{r_{m}^{1} }{\mu} = - \tanh \left( \pi T_\mt{JT} \, t  - \tanh^{-1} (\mu/r_c)  \right) \, , 
\nonumber\\
 x_{m}^{2} &\equiv&\frac{r_{m}^{2} }{\mu} =\ \    \tanh \left( \pi T_\mt{JT} \, t  + \tanh^{-1} (\mu/r_c)  \right) \, .
 \label{lizard1}
\eeqa
Then we write the blackening factor as
\begin{equation}
f(r) = \frac{\mu^2}{L_2^2}\, \tilde f (r/\mu)\qquad \text{where} \quad \tilde f (x) \equiv x^2 -1  \,  .
\end{equation}
Finally we can rewrite eq.~\eqref{Complexity_JT} as
\begin{equation}\label{Complexity_JT_dimless}
\f{d{\cal C}^{\mt{JT}}_{A}}{dt}= -\f{M_{\mt{JT}} }{\pi} \left[ \tilde f(x)\, \log\!\left(     \frac{\ell_{ct}^2}{ L_2^2}   \, \frac{\Phi_{b}^{2}}{\Phi_0^2}\, \f{\mu^2}{ r_c^2 }\, | \tilde f(x)|   \right)   \right]^{x_m^1}_{x_m^2}\, .
\end{equation}
Hence apart from the overall factor of the mass, the above expression is a function of the dimensionless ratios, $\Phi_b/\Phi_0$
and 
\begin{equation}
 2 \pi \,\f{T_\mt{JT}}{\mu_c}=\f{\mu}{ r_c}\qquad{\rm where}\quad
\mu_c\equiv \f{r_c}{L_2^2}\,.\label{liza2}
\end{equation}
That is, $\mu_c$ is the conformal breaking scale in the boundary theory. Both of these ratios should be small as the near-extremal and near-horizon limit requires both  $\frac{\Phi_b}{\Phi_0},\,\frac{\mu}{r_c} \ll1 $. Note that the result \reef{Complexity_JT_dimless} also depends of the ratio $\ell_{ct}/ L_2$, which is an ambiguity that arises in defining holographic complexity with the CA proposal \cite{Vad2} -- see also \cite{NullBound,diverg,Vad1}. We show examples of the time evolution of the holographic complexity for different values of the temperature in figure \ref{JTCA}.

As we already saw in eq.~\reef{JT_NoGrowth}, the late-time limit of the complexity growth is zero. Further, we note that this limit is generically approached from below, given the expression in eq.~\eqref{Complexity_JT_dimless}, in contrast to the expectation from higher dimensional black holes. The leading contribution at late times is given by
\begin{equation}\label{LTimeApprJT}
\f{d{\cal C}^{\mt{JT}}_{A}}{dt}=- 32 \, M_{\mt{JT}} \, \frac{ 2\pi\, \mu_c\,T_\mt{JT}}{\mu_c^2 - (2\pi T_\mt{JT})^2} \, e^{-2 \pi T_{\mt{JT}} \, t}\, T_{\mt{JT}} \, t + \mathcal{O} \left( e^{-2 \pi T_{\mt{JT}} \, t} \right) \, .
\end{equation}

The vanishing of the complexity growth rate at late times may seem puzzling for the JT gravity when we consider that this model is supposed to capture the low energy dynamics of the SYK model, which is maximally chaotic and hence would be expected to exhibit nontrivial complexity growth for very long times. However, from the perspective of the dimensional reduction, this feature is not so surprising. The action \eqref{JT_Action} was derived from the four-dimensional action \eqref{TOTaction} with $\gamma=0$, \ie we did not include the Maxwell boundary term \reef{mba0}, and with a purely magnetic solution, \ie $q_e=0$. Recall that in section \reef{sec:RNBHS}, we found that the late time growth rate also vanished in this case, \eg see eq.~\reef{squareB} with $\chi=0$. The result in eq. \eqref{JT_NoGrowth} was our main motivation to revisit the holographic complexity of charged black holes.

\subsection{JT-like Model}

We now turn our attention to another possible two-dimensional theory that is derived from a purely electrically charged black hole in four dimensions. In the latter black holes, the Maxwell field strength \reef{FieldDyon} has a single component $F_{rt}$ and hence the dimensionally reduced theory incorporates a Maxwell potential and the corresponding field strength is a form of maximal rank in two dimensions. In this sense, the two-dimensional theory has a form reminiscent of the Brown-Teitelboim model \cite{BT1,BT2}, with a dynamical cosmological constant controlled by a field strength of maximal rank. A similar two-dimensional action was also studied in in \cite{Trivedi1992,Lechtenfeld,FIS} and more recently in \cite{Trivedi2}. In \cite{Trivedi2}   it was argued to describe the physics of the extended SYK models with complex fermions with conserved charge studied in \cite{Fu:2016vas, Davison:2016ngz}. When we evaluate the complexity growth for the new dimensionally reduced theory, we find nonvanishing complexity growth at late times, as expected from the higher dimensional analysis in section \ref{subsec:CompDyon}.

Consider again the action \eqref{TOTaction} with $\gamma=0$, assuming the metric ansatz \eqref{metric_ansatz} but a purely electric field, \ie $F$ is supported on the $(x^1,x^2)$-plane. In this case, the bulk action \reef{bulk} reduces to the following two-dimensional action 
\begin{equation}\label{action_preJTLike}
\begin{split}
I_\mt{electric}^\mt{2D} &=\frac{1}{4 G_N} \int_\mathcal{M} d^2x \sqrt{-g}\left(\Psi^2 \mathcal{R} + 2 \left(\nabla \Psi \right)^2 - \widetilde U(\Psi^2)\right)-\frac{\pi}{g^2}\int_\mathcal{M} d^2x \sqrt{-g} \Psi^2 F^2\\
&\qquad\qquad+ \frac{1}{2 G_N} \int_{\partial \mathcal{M}} d x  \, \sqrt{|\gamma|} \, n^{\mu} \nabla_{\mu} \Psi^2   \,,
\end{split}
\end{equation}
with the potential
\begin{equation}\label{potentialpreJT}
\widetilde U(\Psi^2)=-2-6\frac{\Psi^2}{L^2}\,.
\end{equation}
We note that it is possible to show that \eqref{action_preJTLike} leads to the same equations of motion as \eqref{action_preJT} after putting the gauge field on-shell \cite{Almheiri:2016fws}. However, we will show that holographic complexity derived with this action using the CA proposal yields a different result from the JT model studied above.

We recall that the four-dimensional geometry \reef{RNBas} is identical for black holes with $(q_e,q_m)=(0,\qt)$
and $(q_e,q_m)=(\qt,0)$. In particular, the throat of an electrically charged extremal black hole is identical to that of a magnetic extremal black hole. Therefore the corresponding two-dimensional geometry and the (constant) dilaton are identical in the present case as in the previous subsection, \ie as described by eqs.~(\ref{constant_PHI0}--\ref{Lambda2Eq}). Of course, the difference is that the electric throat is supported by a Maxwell field proportional to the two-dimensional volume form $\epsilon_{ab}$, \ie
\beq
(F_0)_{ab}=\frac{\sqrt{4\pi}\,g}{\sqrt{ G_N}} \,\f{q_\mt{T,ext}}{\Phi_0}\,\epsilon_{ab}\equiv E_0\, \epsilon_{ab}\,. \label{extra3}
\eeq

Now, as in the previous subsection, we wish to construct a theory which captures small deviations from the extremal throat. Hence, we expand the dilaton as in eq.~\reef{small} with $\Phi/\Phi_0\ll1$. However, in the present case, we also wish to capture small corrections to the extremal field strength in eq.~\reef{extra3}. As a consequence, we also expand the field strength as
\beq
F_{ab}=(F_0)_{ab}+\tilde f_{ab} = 2\, \partial_{[a}(A_0)_{b]}
+2\,\partial_{[a}\tilde a_{b]}\,,
\label{mfield}
\eeq
where $\tilde f$ captures corrections of order $\Phi/\Phi_0$ relative to $F_0$. 
When we expand the bulk action in eq.~\reef{action_preJTLike} to linear order in both $\Phi$ and $\tilde f$, the resulting action takes the form 
\beqa\label{Act_JTlike}
I^{\mt{JT-like}}_{\mt{bulk}} &=& I^{\mt{JT}}_{\mt{bulk}} +\frac{ E_0^2}{2g^2}\int_{\mathcal{M}}d^2x \sqrt{-g} \left[\Phi_0-\Phi\right]\\
&&\quad -\frac{1}{4g^2}\int_{\mathcal{M}}d^2x \sqrt{-g}\left[(\Phi_0+\Phi) (F_0)^2 + 2 \Phi_0\, (F_0)^{ab} \tilde f_{ab} \right] \, .
\nonumber
\eeqa
where $I^{\mt{JT}}_{\mt{bulk}}$ is precisely the action given in eq.~\reef{JT_Action}. We will call this theory the ``JT-like'' model. 
As before, $\Phi_0$ is simply treated as a constant parameter defining the theory, as is the constant $E_0$ defined in eq.~\reef{extra3}. In contrast, we treat $A_0$ as a dynamical field, however, our prescription is that the solution is always chosen to yield precisely the extremal field strength in eq.~\reef{extra3}. Again, the deviations from this extremal form are captured by $\tilde f$.  It is useful to write out the full equations of motion
\beqa
\delta \tilde a_a\ :&&\quad 0=\nabla_a (F_0)^{ab}\,,
\label{two}\\
\delta \Phi\ :&&\quad 0={\cal R}-2\Lambda_2 -\f{4\pi G_N}{g^2}\left((F_0)^2 + 2 E_0^2\right)\,,
\label{one}\\
\delta (A_0)_a\ :&&\quad 0=\nabla_a \tilde f^{ab}+ \f{\nabla_a\Phi}{\ \Phi_0}\, (F_0)^{ab}\,,
\label{three}\\
\delta g_{ab}\ :&&\quad 0= -\nabla_a\nabla_b\Phi+g_{ab}\left(\nabla^2\Phi+\Lambda_2\Phi
\right)
\nonumber\\
&&\qquad\qquad
-\f{2\pi G_N}{g^2}\,\Phi_0\left(4(F_0)_a{}^c(F_0)_{bc}+g_{ab} \left(2E_0^2-(F_0)^2 \right)\right)
\label{four}\\
&&\qquad\qquad
-\f{2\pi G_N}{g^2}\,\Phi\left(4(F_0)_a{}^c(F_0)_{bc}-g_{ab} \left(2E_0^2+(F_0)^2 \right)\right)\nonumber\\
&&\qquad\qquad
-\f{2\pi G_N}{g^2}\,\Phi_0\left(4(F_0)_a{}^c\tilde f_{bc}
+4\tilde f_a{}^c(F_0)_{bc}-2g_{ab} (F_0)^{cd}\tilde f_{cd} \right)\nonumber
\eeqa
Of course, eq.~\reef{two} yields the solution $(F_0)_{ab}\propto \epsilon_{ab}$ and again, our prescription is that we should choose the prefactor to be the extremal electric field $E_0$, defined in eq.~\reef{extra3}. When we set $F_0$ to its extremal value, we note that $(F_0)^2 =-2 (E_0)^2$ and eq.~\reef{one} reduces to the expected $0={\cal R}-2\Lambda_2$. That is, the two-dimensional geometry becomes locally AdS$_2$ with a curvature set by $\Lambda_2$ and hence we may write the solution as in eq.~\reef{AdScoord} for the JT model. Further, the mass, entropy and temperature all take the same form as given in eq.~\reef{MSTJT}.

Now in eq.~\reef{three}, we have dropped two terms proportional to $\nabla_a (F_0)^{ab}$ since this factor already vanishes according to eq.~\reef{two}. We can write the solution of eq.~\reef{three} as
\beq
\tilde f_{ab}= \left[\delta E -\f{\Phi}{\Phi_0}\,E_0 \right] \epsilon_{ab}\,.
\label{five}
\eeq
The first term represents a small shift in the background electric field,\footnote{One might also think of this as a shift in the extremal charge, with $\delta q_\mt{T}= q_\mt{T,ext}\,\delta E/E_0$. \label{foot99}} which is allowed by the dynamical Maxwell field --- note that we assume $\delta E/E_0\lesssim \Phi_b/\Phi_0$. The second term represents the leading correction to the field strength created by the running of the dilaton. 

Lastly, we turn to the dilaton equation of motion in eq.~\reef{four}. We note that the second line is actually the leading contribution since it is not suppressed by a factor of $\Phi$ or $\tilde f$. However, when we substitute $(F_0)_{ab}=E_0\epsilon_{ab}$, this collection of terms vanishes. Upon substituting this extremal field as well as the perturbation \reef{five}, eq.~\reef{four} reduces to 
\beq
0= -\nabla_a\nabla_b\Phi+g_{ab}\left(\nabla^2\Phi+\Lambda_2\Phi
\right)
+\f{8\pi G_N}{g^2}\,\Phi_0\,E_0\,\delta E\, g_{ab} \,.\label{six}
\eeq
We can absorb the last term with a simple constant shift of the dilaton, \ie 
\beq
\tilde\Phi = \Phi + \Phi_q\qquad
{\rm where}\quad \Phi_q\equiv\f{8\pi G_N}{g^2}\,\f{E_0\Phi_0}{\Lambda_2}\,\delta E \,,
\label{seven}
\eeq
Then $\tilde\Phi$ satisfies the dilaton equation appearing for the JT model. Hence our final dilaton solution for the JT-like model becomes
\beq
\Phi =  \Phi_q\left(\f{r}{r_c}-1
\right)+\Phi_b\, \f{r}{r_c}\,.
\label{eight}
\eeq
The parameters are chosen so that $\Phi_b$ again corresponds to the value of the dilaton at the boundary $r=r_c$. However, as a result, the dilaton has a new value when evaluated at the horizon $r=\mu$ and so the entropy \reef{MSTJT} is shifted by a small amount proportional to $\delta\qt$ (relative to the JT model).

As described above the geometry is precisely the same as in the JT model and so the Penrose diagram in figure \ref{WDWpatch} still describes the solution for the JT-like model \reef{Act_JTlike}. The new features are a small  shift of the dilaton proportional to $\delta\qt$ in eq.~\reef{eight} (\ie compared to the solution \reef{AdScoord} for the JT model), and the field strengths $F_0$ and $\tilde f$ which capture the extremal Maxwell field and the leading correction to this extremal two-form.

\subsubsection*{Complexity Growth}

\begin{figure}
\centering
\includegraphics[height=6cm]{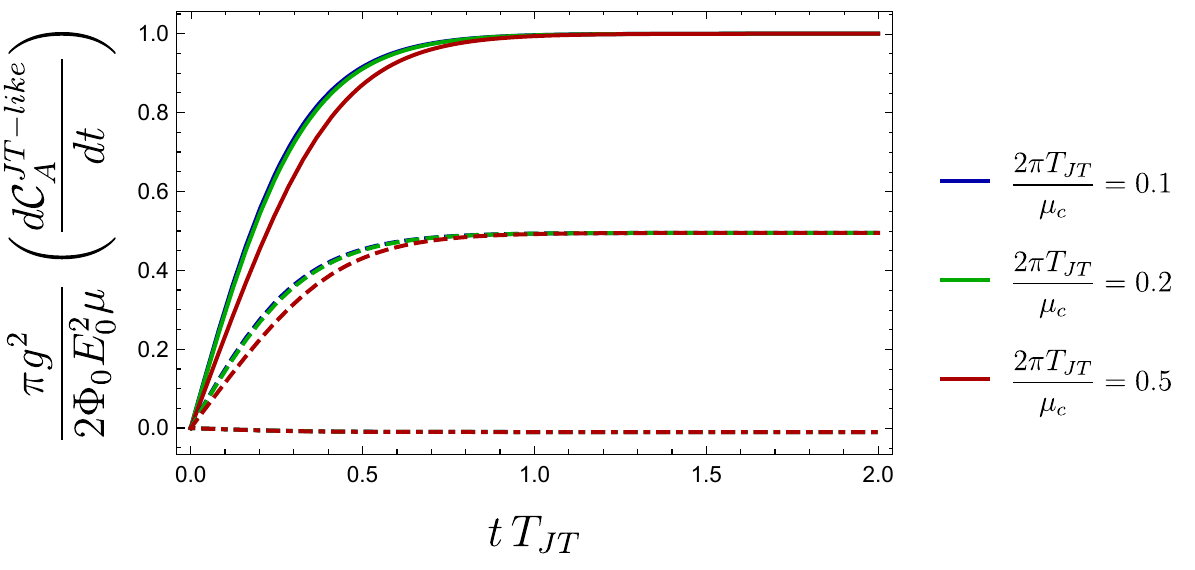}
\caption{Complexity growth in the dimensionally reduced model derived from the RN black holes with $q_{m}=0$, the ``JT-like'' model given by the action in eq.~\eqref{Act_JTlike}. We fix the dimensionless ratios $\Phi_b = 10^{-3} \Phi_0$,  $4 \pi L_2^2 = 10^{-4} \Phi_0$, $\delta E = 0.01 E_0$ and $l_{\mt{ct}} = L_2$. The solid curves are the complexity growth without the addition of the counterterm $\gamma=0$, the dashed curves correspond to $\gamma=\frac{1}{2}$ and the dot-dashed correspond to $\gamma= 1 $ . As in the electrically charged black holes discussed in section \ref{sec:RNBHS}, a higher value of the parameter $\gamma$ decreases the late time growth of complexity for the ``JT-like model''.}
\label{JTlikeCA}
\end{figure}

Given the close connection of eq.~\reef{Act_JTlike} to the JT action \reef{JT_Action}, we can express the on-shell bulk action as
\begin{equation}\label{JTlikeOnshell}
I^{\mt{JT-like}}_{\mt{bulk}} |_\mt{on-shell} = I^{\mt{JT}}_{\mt{bulk}} |_\mt{on-shell}+ \frac{\Phi_0\, E_0^2}{g^2}\int_{\mathcal{M}}d^2x \sqrt{-g}\left(1+\frac{\delta E}{E_0}-\frac{\Phi}{\Phi_0}\right) \, ,
\end{equation}
Further, while this expression only refers to the bulk action, it is clear that the surface terms for the null boundaries of the WDW patch are dimensionally reduced in exactly the same way as before, \ie the null joint terms and the null surface counterterm are given by eq.~\reef{JTjntct}. Hence we can easily extend the analysis of the holographic complexity for the JT model from the previous subsection by simply investigating the contribution of the second term in  eq.~\reef{JTlikeOnshell} to the CA proposal \reef{defineCA}.  In fact, we found that the late-time growth rate of the holographic complexity vanished for the JT model, and so in the JT-like model, the late-time growth rate will come entirely from the time derivative of this term,
\beqa\label{JTlikeApprox}
&&\frac{d}{dt} \left[\frac{\Phi_0\, E_0^2}{g^2}\int_{\mathcal{M}}d^2x \sqrt{-g}\left(1+\frac{\delta E}{E_0}-\frac{\Phi}{\Phi_0}\right)  \right]\\
&&\qquad\qquad=\frac{\Phi_0\, E_0^2}{g^2}\left[\left(1+\frac{\delta E}{E_0}+\f{\Phi_q}{\Phi_0}\right) r-\frac{\Phi_q+\Phi_b}{2\Phi_0 r_c}\,r^2 \right]_{r_m^1}^{r_m^2}\,.\nonumber
\eeqa
Again at late times, we have $r_m^1\to -\mu$ and $r_m^2\to \mu$ and therefore the complexity growth rate becomes\footnote{Given the prefactor in eq.~\reef{seven}, we note that
\beq
\f{\Phi_q}{\Phi_0}= - 2  \left( \frac{L^2 + 3 r_h^2}{L^2 + 6 r_h^2} \right) \,\f{\delta E}{E_0}
\eeq 
and hence both of the corrections are the same order in the second factor of eq.~\reef{hip5}. In fact for large black holes, \ie $r_h/L \gg1$, these two corrections will cancel one another.} 
\begin{equation}
\lim_{t \to \infty}\frac{d \mathcal{C_A^{\mt{JT-like}}}}{dt}=
\frac{2\Phi_0\, E_0^2\,\mu}{\pi g^2}\left(1+\frac{\delta E}{E_0}+ \f{\Phi_q}{\Phi_0}\right)  
\,.
\label{hip5}
\end{equation}
for the JT-like model.  Again, the nonvanishing result here may not be so surprising since the corresponding (electrically charged) black holes in four dimensions also exhibited a constant growth rate at late times. In fact, a careful translation of the parameters shows that eq.~\reef{hip5} matches the higher dimensional result in eq.~\reef{square} to leading order in the near extremal limit.\footnote{That is, we substitute $r^\mt{4D}_{\pm} = r_h \pm \mu$ and and $(q_e,q_m)=(q_\mt{T,ext}+\delta q_\mt{T},0)$ (with $\delta q_\mt{T}$ from footnote \ref{foot99}) in eq.~\reef{square} and expand to linear order in both $\mu$ and $\delta q_\mt{T}$. Then the leading terms for the late-time growth in eqs.~\reef{square} and \reef{hip5} agree, \ie the $O(\mu)$ terms agree, the $O(\delta q_\mt{T})$ terms agree in that they vanish, but the $O(\mu\,\delta q_\mt{T})$ terms disagree. \label{foot88}} The expression in eq.~\reef{JTlikeApprox} (divided by $\pi$) can be combined with eq.~\reef{Complexity_JT} to give a full description of the growth rate for the holographic complexity in the JT-like model. Some examples of the full time profile of the growth rate are illustrated in figure \ref{JTlikeCA}. 

\subsection{Boundary Terms?}

For both the JT and the JT-like model, we found that the growth of the holographic complexity matched (at least qualitatively) the results found in section~\ref{subsec:CompDyon} for the corresponding black holes in four dimensions. In section \ref{sec:MBA}, we also found that the Maxwell surface term \reef{mba0} can have a dramatic effect on the growth rate and so in the following, we investigate the dimensional reduction of this surface term and its effect on the holographic complexity in the two-dimensional gravity theories.

\subsubsection*{JT-like model}

We start by analyzing the role of the Maxwell surface term in the JT-like model. In this case, the Maxwell field is still a dynamical field in the two-dimensional theory and so eq.~\eqref{mba0} reduces in a straightforward way to a boundary term for the WDW patch in two dimensions. Substituting the ansatz \eqref{metric_ansatz}, the two-dimensional surface term becomes
\begin{equation}
I^{\mt{2D,elec}}_{\mu \mt{Q}}=\frac{4 \pi \gamma}{g^2}\int_{\partial \mathcal{M}} d\Sigma_a F^{ab} A_b \Psi^2\,.
\label{elec02}
\end{equation}
Expanding to linear order in the dilaton with eq.~\eqref{small} and in the Maxwell perturbation $\tilde a_a$ in eq.~\reef{mfield}, we find 
\begin{equation}
I^{\mt{JT-like}}_{\mu \mt{Q}} = \frac{\gamma}{g^2}\int_{\partial \mathcal{M}} d\Sigma_a \left((F_0)^{ab} (A_0)_b (\Phi_0+\Phi) +\Phi_0\left((F_0)^{ab}\, \tilde a_b+\tilde f^{ab} (A_0)_b\right) \right)\,.
\label{elec03}
\end{equation}

Alternatively, when the Maxwell field is on-shell, we can also use eq.~\reef{MaxBdryActOnS} to express the Maxwell surface term in terms of a bulk integral. The two-dimensional version of this bulk integral becomes
\begin{equation}
I^{\mt{2D,elec}}_{\mu \mt{Q}} \big{|}_\mt{on-shell} = \frac{2\pi \gamma}{g^2}\int_{\mathcal{M}} d^2 x \, \sqrt{-g} \,\Psi^2\,  F^{ab} F_{ab} \,,
\label{elec04}
\end{equation}
or after the usual linear expansion, we arrive at
\begin{equation}
I^{\mt{JT-like}}_{\mu \mt{Q}} \big{|}_\mt{on-shell} = \frac{ \gamma}{2g^2}\int_{\mathcal{M}} d^2 x \, \sqrt{-g} \left( (F_0)^{ab} (F_0)_{ab}\, (\Phi_0+\Phi) + 2\Phi_0\, (F_0)^{ab} \tilde f_{ab}
\right)\,.
\label{elec05}
\end{equation}

Next, we evaluate eq.~\reef{elec03} (or equivalently eq.~\reef{elec05}) on the WDW patch, with the solution given by the metric \reef{AdScoord}, the dilaton \reef{eight}, the extremal field strength \reef{extra3} and the perturbation \reef{five} to the Maxwell field. The time derivative then yields 
\beq \label{elec06}
\f{d I^{\mt{JT-like}}_{\mu \mt{Q}} }{dt} = - \frac{\gamma \, E_0^2}{g^2} \left[ r \left( \Phi_0 + 2 \Phi_0\,\frac{ \delta E }{E_0} + \Phi_q  \right)  - \frac{r^2}{2 r_c} \left( \Phi_b + \Phi_q \right) \right]  \bigg{|}_{r^1_m}^{r^2_m}\,.
\end{equation}
Without the Maxwell surface term, the late-time growth rate for the JT-like model was given by eq.~\reef{hip5} and so combining this result with the contribution from the Maxwell surface term \reef{elec03} then yields
\begin{equation}
\lim_{t \to \infty}\frac{d \mathcal{C_A^{\mt{JT-like}}}}{dt}=
\frac{2\Phi_0\, E_0^2\,\mu}{\pi g^2}\left[ \left(1 -\gamma \right) \left( 1+ \f{\Phi_q}{\Phi_0}  \right) + \left( 1- 2 \gamma \right) \frac{\delta E}{E_0} \right]  \,.
\label{hip5a}
\end{equation}
Comparing this expression to the late-time growth for four-dimensional black holes in section \ref{sec:MBA}, we find agreement between eqs.~\reef{square2} and \reef{hip5a} to leading order in the near-extremal limit --- see footnote \ref{foot88}. 

\subsubsection*{JT model}
As we described in section \ref{wonderland}, the JT model arises from the dimensional reduction of a magnetically charged black hole and in this case, the Maxwell field is completely ``integrated out" in the dimensional reduction. Further, as we explained in Appendix \ref{MBAapp}, the evaluation of the Maxwell surface term \eqref{mba0} for the magnetically charged black holes is more subtle. The interesting contributions to the surface term actually live on the surface(s) dividing the patches where the gauge potential is well defined, \eg see eq.~\reef{poult5}. Alternatively, we can again use the bulk expression in eq.~\reef{MaxBdryActOnS} for the on-shell Maxwell field, which yields precisely the same result as shown in eq.~\eqref{poult3}.  Expressing the former in terms of the two-dimensional variables, the surface term becomes 
\begin{equation} \label{magnetic_reduction}
I^{\mt{2D,mag}}_{\mu \mt{Q}}=\frac{\gamma q_{m}^2}{G_N}\int_\mathcal{M}d^2x \sqrt{-g}\,\frac{1}{\Psi^2}\,.
\end{equation}
Then substituting eq.~\reef{small} and expand to first order in $\Phi$, we see that the Maxwell surface term contributes as
\beq \label{mag2}
I^{\mt{JT}}_{\mu \mt{Q}}=\frac{4\pi \gamma q_\mt{T,ext}^2}{\Phi_0 G_N}\int_\mathcal{M}d^2x \sqrt{-g}\,\left(1-\f{\Phi}{\Phi_0}\right)\,,
\end{equation}
in the JT model. Recall that this two-dimensional model describes gravity in the near-extremal throat with $q_m=q_\mt{T,ext}$. Note that since the magnetic Maxwell field and the relevant surfaces are integrated out in the dimensional reduction, there is no way to think of eqs.~\reef{magnetic_reduction} or \reef{mag2} as a surface term in the two-dimensional theory. Rather, here we are modifying the standard CA prescription in the JT model by adding a new bulk contribution to the holographic complexity. In particular, we observe that the first contribution in eq.~\reef{mag2} is simply proportional to the spacetime volume of the WDW patch and so this contribution is reminiscent of the CV2.0 proposal \cite{fish2} where the holographic complexity is equated with the spacetime volume to the WDW patch. That is, when the Maxwell surface term is included in the usual complexity=action prescription in four dimensions, the dimensional reduction produces a complexity=(action +  spacetime volume) prescription for the JT model. 

Evaluating the above expression 
\reef{mag2} on the WDW patch with the solution \reef{AdScoord} and considering the time derivative then yields 
\beq \label{mag3}
\f{d I^{\mt{JT}}_{\mu \mt{Q}} }{dt} =\frac{4\pi \gamma q_{m}^2}{\Phi_0 G_N} \left(r - \frac{\Phi_b \, r^2 }{2 \Phi_0 r_c} \right) \bigg{|}_{r^1_m}^{r^2_m}\,.
\end{equation}
Without the Maxwell surface term, the late-time growth rate for the holographic complexity vanished, as shown in eq.~\reef{JT_NoGrowth}. Hence once we reconsider the holographic complexity with the addition of the ``surface" term \reef{mag2}, the late-time growth rate is governed entirely by eq.~\reef{mag3} which yields
\begin{equation}\label{JT_Growing}
 \lim_{t \to \infty}\frac{d \mathcal{C}^{\rm JT}_{A}}{d t}   = \frac{8 \gamma q_{m}^2  \mu}{\Phi_0 G_N} \, .
\end{equation}
Comparing this result to those in four dimensions, we see that this new expression matches to linear order in $\mu$ the rate in eq.~\eqref{square2} with $(q_e,q_m)=(0,q_\mt{T,ext})$, as well as $r^\mt{4D}_{\pm} = r_h \pm \mu$.

\section{Discussion}\label{sec:Discussion}

We began in section \ref{sec:RNBHS}, by investigating the complexity$=$action proposal \reef{defineCA} on the holographic complexity of charged four-dimensional black holes in the usual Einstein-Maxwell theory \reef{bulk}. We found that the results were very sensitive both to the type of charge (\ie electric versus magnetic) and to the inclusion of the Maxwell boundary term \reef{mba0}. Without the latter surface term (\ie $\gamma=0$), the late-time growth rate vanished for black holes carrying purely magnetic charge, while for the electrically charged case, it is a nonvanishing constant in accord with the general expectations of eq.~\reef{general}. The general result for dyonic black holes carrying both kinds of charge is given in eq.~\reef{squareB}. In section \ref{subsec:Shocks}, we also noted that the switchback effect exhibited a similar sensitivity to the type of charge.

However, this picture changes dramatically when the Maxwell boundary is included. For example, with $\gamma=1$, the roles of the electric and magnetic charges described above are reversed, \ie the late-time growth rate vanishes with electric charge and is nonvanishing with magnetic charge. The behaviour of the late-time growth for general $\gamma$ is given in eq.~\reef{square2}. In particular, we found that the electric and magnetic charges contribute on an equal footing with the choice $\gamma=1/2$. We might recall that when we are evaluating the Maxwell boundary term for an on-shell gauge field that we can express the contribution as a bulk integral of $F_{\mu\nu}F^{\mu\nu}$, \ie with the same form as the bulk Maxwell action \reef{bulk}. Hence with $\gamma=1/2$, these boundary and bulk contributions precisely cancel, as is evident in eq.~\reef{combo}. A possible alternative then would be to define complexity=(gravitational action). That is, we could drop both the bulk and boundary terms involving the Maxwell field for eq.~\reef{TOTaction} to define the ``gravitational action" (\ie we only keep the geometric contributions to the action) and then define the complexity by evaluating this action on the WDW patch. Of course, because the charges only enter the metric \reef{RNBas} through the combination $\qt^2= q_e^2 + q_m^2$, the complexity only depends on the same combination with this definition. It might be interesting to investigate this proposal in other settings.

In section \ref{sec:Dila}, we investigated the CA proposal \reef{defineCA} for charged black holes in a family of Einstein-Maxwell-Dilaton theories \reef{pda}.  As the value of the parameter $\alpha$ (controlling the coupling between the dilaton and the Maxwell field) is varied, the nature of the curvature singularities and the causal structure of the black holes change, as shown in figure \ref{CausalDilaton}. Generally, the late-time growth rate of the holographic complexity was nonvanishing for the electrically charged black holes. However, when the Maxwell boundary term \eqref{mbadila} was included, we showed in eq.~\reef{DilaNewmann} that choosing $\gamma_\alpha=1$ sets the late-time growth rate to zero for the cases where the causal structure was like that of the Reissner-Nordstrom-AdS black holes, \ie for $\alpha^2<1/3$. No such choice was possible when the causal structure appeared as in the Schwarzschild-AdS black holes, \ie for $\alpha^2\ge 1/3$. The former behaviour was analogous to that found in the Einstein-Maxwell theory and therefore it appears that the causal structure of the black hole was one of the essential features producing the unusual behaviour found in section \ref{sec:RNBHS}. 

Let us note, however, that the analysis in section \ref{sec:Dila} did not consider dyonic or magnetically charged black holes for the simple reason that, to the best of our knowledge, such solutions have not yet been constructed for the Einstein-Maxwell-Dilaton theory \reef{pda}. In the Einstein-Maxwell theory \reef{bulk}, it is straightforward to produce magnetic solutions given the electrically charged black holes using electric-magnetic duality. This operation is not as straightforward for the Einstein-Maxwell-Dilaton theories, in which case it is natural to replace
\beq
F_{\mu\nu}\to \widetilde F_{\mu\nu}=\frac{e^{-\alpha\phi}}2\, \varepsilon_{\mu\nu\rho\sigma} F^{\rho\sigma}\,,
\qquad \phi \to \tilde \phi =-\phi\,.
\label{dualEMD}
\eeq
The action of the ``dual" theory is then
\begin{equation}\label{pdaXX}
I_\mt{bulk} = \ \frac{1}{16 \pi G_N} \int_\mathcal{M} d^{4} x \sqrt{-g} \left(\mathcal{R}-2(\partial \tilde\phi )^2-V(-\tilde\phi)\right)-\frac{1}{4g^2}\int_{\mathcal{M}} d^{4}x \sqrt{-g}e^{-2\alpha\tilde\phi}\widetilde F_{\mu \nu}\widetilde F^{\mu \nu}\,,
\end{equation}
which matches eq.~\reef{pda} except for the appearance of $V(-\tilde\phi)$. Generally then, with the transformation \reef{dualEMD}, the electrically charged solutions \reef{metric_dilaton}  of the original theory would become magnetically charged solutions of the new theory \reef{pdaXX}. However, as we noted above eq.~\reef{metric_dilaton}, there are three special cases for which $V(-\phi)=V(\phi)$ and hence for which eq.~\reef{dualEMD} leaves the theory invariant. Note that all three of these special cases, \ie $\alpha^2=1/3,\ 1$ and 3, lie in the regime where the causal structure matches that of the Schwarzschild-AdS black hole. Hence at least of these three cases, it is straightforward to verify that the late-time growth of the magnetic black holes is nonvanishing. It would, of course, be interesting to construct magnetic or dyonic black holes for the Einstein-Maxwell-Dilaton theories \reef{pda} more generally and to fully investigate holographic complexity in these theories.

In section \ref{sec:JT}, we turned to holographic complexity for black hole in two dimensions. In particular, we showed that the late-time growth rate vanishes for the JT model in eq.~\reef{JT_NoGrowth}. The latter mirrored the behaviour for the magnetic Reissner-Nordstrom-AdS black holes in four dimensions, which play a role in constructing the JT action \reef{JT_Action} via dimensional reduction. This situation can be ameliorated by instead considering a dimensional reduction describing the near-horizon physics of near-extremal electric black holes (in four-dimensions). In the resulting JT-like theory \reef{Act_JTlike}, the late-time growth rate is nonvanishing, as shown in eq.~\reef{hip5}. We also considered the dimensional reduction of the Maxwell boundary term \reef{mba0} in both cases. Again, the effect of this surface term mimicked that found in four dimensions. In particular, the late-time growth rate of the JT model is now nonvanishing, as shown in eq.~\reef{JT_Growing}. The slightly unusual feature is that as a result of including the Maxwell boundary term in four dimensions, the complexity=action prescription becomes a complexity=(action +  spacetime volume) prescription for the JT model, \ie we have a mixture of the CA and CV2.0 proposals in two dimensions. We will discuss the two-dimensional results further below.

\subsubsection*{Maxwell Boundary Term, Revisited}

In section \ref{sec:RNBHS}, we found that the Maxwell boundary term played an essential role in order for the CA proposal to produce the expected properties of the complexity (\eg late-time growth and the switchback effect) for dyonic black holes carrying both electric and magnetic charges. A priori, this surface term would not appear to be an essential part of the Einstein-Maxwell action, \eg obviously, it does not affect the equations of motion.  It should not be surprising that holographic complexity, or more specifically the CA proposal, can be sensitive to surface terms since an analogous behaviour was already observed with the null counterterm \reef{Counterterm} in the gravitational action \cite{Vad1,Vad2}. In particular, in shock wave geometries, this surface term plays an essential role in ensuring that the holographic complexity exhibits both late-time growth and the switchback effect. Of course, another guiding principle that suggests that this boundary term should be included is to ensure that the action is invariant under reparametrizations of the null boundaries, as emphasized in \cite{NullBound,Reynolds:2016rvl}. In fact, it is this principle that fixes the overall coefficient with which the null counterterm is added to the action. In the present case, we have not definitely fixed the coefficient of the Maxwell boundary term. We did find that $\gamma= 1/2$ seems to be special as it allows both magnetic and electric charges to participate in``computations'' on an equal footing, as can be seen from eq.~\eqref{square2}. Hence it may be that electric-magnetic duality (or S-duality) provides the guiding principle to fix $\gamma$. However, it is not immediately obvious (to us) that this is the correct choice --- see further comments below. 

While Maxwell boundary term does not effect the equations of motion, it does play a role in the variational principle by changing the boundary conditions imposed on the gauge field, as described around eq.~\reef{mixed}. As we noted previously, in this way, this surface term plays a role in black hole thermodynamics, \ie it becomes an important part of the Euclidean action depending on the thermodynamic ensemble of interest  \cite{char0, char1,char2}. Hence one might ask if different ensembles will ``compute'' differently, \ie if the electric and magnetic charges would make distinct contributions in different ensembles.\footnote{Implicitly, we are suggesting that the action used to evaluate the holographic complexity \reef{defineCA} would be the same as that used to evaluate the Euclidean action for the thermodynamic ensemble.} Preliminary investigations with simple qubit models seem to indicate that this is indeed the case \cite{beni18}, but of course, it would be interesting to study this question further.

Even though the Maxwell term is associated with modifying the boundary conditions for the field equations, we are not suggesting that these boundary conditions should be applied on the boundary of the WDW patch. For example, irrespective of the choice of $\gamma$, if we are evaluating the holographic complexity with CA prescription for a boundary state dual to a solution in which various charges are circulating in the bulk spacetime, the bulk solution should not be modified and the charges would appear to freely flow into or out of the WDW patch, as discussed in \cite{Brown1,Brown2}.  On the other hand, we might wonder if  ``quantum'' (\ie finite-$N$) corrections to this saddle point evaluation of the holographic complexity would involve fluctuations of fields on the WDW patch which respect the boundary conditions determined by our choice of surface terms. 

In passing, we note that while the holographic complexity defined by the CA proposal is sensitive to the presence of the Maxwell boundary term, it is only the infrared properties which exhibit this sensitivity.
The asymptotic contributions to action are essentially independent of the choice of the coefficient $\gamma$. For example, for an electrically charged black hole given in eqs.~\reef{RNBas} and \reef{FieldDyon}, we find 
\begin{align}
I_{\mu Q} (\text{UV}) & \sim \frac{4 \gamma\, q_e^2}{G_N\,r_+}   \int^{r_{\text{max}}} \frac{d r}{f_{\RNo}(r)} \ + \cdots \nonumber \\
&\sim - \frac{4\gamma\, L^2\,q_e^2}{ G_N\,r_+\,r_{\text{max}}}  + \cdots \, .
\end{align}
Hence the leading UV contribution vanishes in the limit $r_{\text{max}} \rightarrow \infty$, \ie there are no UV divergent contributions coming from the Maxwell boundary term. Therefore the holographic complexity has the same UV structure \cite{diverg}, independent of the choice of $\gamma$ in the boundary term. 

However, to close our discussion here, we observe that the limit of zero charge is subtle. Naively, from perspective of the boundary CFT, nothing particularly strange should happen in the limit $\qt \to 0$. On the other hand, from the bulk perspective, there is a nontrivial and abrupt change in the causal structure of spacetime with $\qt \to 0$. If we consider this limit for the late-time growth rate \reef{square2}, we find
\begin{equation}
 \lim_{t \to \infty}\frac{d \mathcal{C}_A}{d t}\bigg{|}_{\qt\rightarrow 0}   =  \frac{2 M }{\pi}\,\frac{(1-\ap)\chi^2+\ap}{1+\chi^2}\, . \label{square2X}
\end{equation}
The first factor corresponds to the late time growth rate of a neutral black hole \cite{Brown1,Brown2}, and hence we only recover this expected rate when the second factor is equal to one. That is, for each choice of the coefficient $\gamma$, there will only be one class of black holes, \ie with $\chi^2 = (\gamma - 1)/\gamma$, for which the expected rate is recovered in the zero-charge limit.\footnote{Notice that we can only produce the desired limit with a phyiscal charge ratio (\ie $\chi^2\ge0$) for either $\gamma\ge1$ or $\gamma\le0$.} We do not have any insight into this issue, but let us add that similar subtleties arise in the zero-charge limit for the Einstein-Maxwell-Dilaton theory studied in section \ref{sec:Dila}, and also with higher curvature corrections \cite{Love2}.

\subsubsection*{Back to Two Dimensions}

In section \ref{sec:Dila}, the causal structure was identified as an essential feature in determining the unusual behaviour of the holographic complexity for charged black holes. Therefore, since the AdS$_2$ black holes \reef{AdScoord} inherit a causal structure with an outer and inner horizon from the near-extremal black holes in higher dimensions \cite{Carroll:2009maa}, it is not surprising that the holographic complexity exhibits behaviour analogous to that found for the four-dimensional charged black holes in section \ref{sec:RNBHS}. For example, the vanishing of the late-time growth rate for the CA proposal found for the JT model in eq.~\reef{JT_NoGrowth} matches with the vanishing rate found for the four-dimensional black holes carrying only magnetic charge (and using $\gamma=0$) in eq.~\reef{square}.

On the other hand, the close parallels between the two- and four-dimensional results may seem unexpected when one recalls that the dimensional reduction producing the JT and JT-like models focuses on the near-horizon region of near-extremal black holes in four dimensions. That is, the cut-off surface at $r=r_c$ introduced for the two-dimensional black holes in figure \ref{WDWpatch} is implicitly a constant radius surface deep in the interior of the corresponding four-dimensional solution. Hence the WDW patches correspond to very different regions of the spacetime in the two different contexts. That is, the four-dimensional WDW patch is anchored to a cut-off surface near the asymptotic AdS$_4$ boundary, while the two-dimensional WDW patch is anchored to a constant radius surface deep in the throat of the four-dimensional black hole. 

Given these differences, one must examine the results more closely to understand the similarities in the complexity growth rates in two and four dimensions. First, in eqs.~\reef{CARNdyo} and \reef{BtDer2}, we see that all of the contributions to the four-dimensional growth rate correspond to terms evaluated at the meeting junctions at $r=r^{1,2}_m$. Of course, both of these junctions are in the throat and at late times, they approach the inner and outer horizons, \ie $r^1_m\to r_-$ and $r^2_m\to r_+$. That is, the growth rate is  determined ``infrared" part of the four-dimensional geometry, \ie by the near-AdS$_2$ throat of the near-extremal black holes. Further, in these geometries with two horizons, the late-time rate is completely determined by quantities evaluated at the corresponding bifurcation surfaces. Of course, we find the same behaviour for the contributions to the growth rate in the two-dimensional geometry, \eg see eqs.~\reef{Complexity_JT} and \reef{JTlikeApprox}, and this explains the close match between the results in two and four dimensions.\footnote{We should add that the dimensional reduction does not require any modification of the time coordinate and hence another important ingredient in this match is that the rates in two and four dimensions are measured with respect to the same time coordinate \cite{HolographicJT}.}

In fact, one finds that the same late-time growth rate will be derived for WDW patches anchored to any fixed $r$ surfaces in the four-dimensional geometry \cite{Leon00}. This again points to the importance of the causal structure in determining the behaviour of the holographic complexity for the CA proposals. However, we note that when the cut-off or anchor radius is varied, the details of the early-time transients are modified. Further, let us add that there have been significant developments concerning the holographic interpretation of moving the AdS boundary into the bulk as a $T\bar{T}$ deformation of the boundary theory \cite{Zamolodchikov:2004ce,Smirnov:2016lqw, McGough:2016lol, Kraus:2018xrn, Taylor:2018xcy, Hartman:2018tkw,Donnelly:2018bef}. Some recent progress in connecting these developments with holographic complexity was reported in \cite{Akhavan:2018wla},\footnote{We note that refs.~\cite{Alishahiha:2018swh, Akhavan:2018wla} suggest a very different understanding of holographic complexity for JT gravity than developed here (and in \cite{Brown:2018bms}). In particular, this approach relies on defining a new cut-off surface behind the horizon using the Lloyd bound \cite{Lloyd}. However, we must remind the reader that the conjectured relation \cite{Brown1,Brown2} between the CA proposal and the Lloyd bound is known to fail \cite{Growth, fish, brian, Alishahiha:2018tep}.} however, it remains an interesting future direction to fully develop these connections.

One of our most interesting results in section \ref{sec:JT} was that when the Maxwell surface term is included in the usual complexity=action prescription in four dimensions, the dimensional reduction produces a complexity=(action +  spacetime volume) prescription for the JT model.\footnote{We observe that in this construction, the spacetime volume contribution \reef{mag2} comes with a very specific prefactor. In general, the normalization of the holographic complexity is an ambiguity for the CV2.0 prescription \cite{fish2}.}
Note that with the latter prescription, the purely ``topological" sector (representing extremal black holes with $\Phi=0$) exhibits complexity growth. Of course, this is in keeping with the recent results of \cite{Brown:2018kvn} indicating that the entire entropy of the extremal black hole contributes to the late-time complexity growth \reef{general}. This also reminds one of the suggestion that the JT model should be interpreted as a low-energy sector embedded in a quantum gravity theory with a larger Hilbert space \cite{twodans}. Of course, the latter is a natural perspective here where the two-dimensional model and the prescription for holographic complexity were both derived with a dimensional reduction from four dimensions.

\subsubsection*{Other Future Directions}

We have already commented on various future directions above, but let us close with a few more observations. The preceding discussion of the JT model reminds us of the other prescriptions of holographic complexity, in particular, the CV proposal \reef{defineCV} and the CV2.0 proposal. While our analysis in this paper focused almost entirely on the CA proposal \reef{defineCA}, it is straightforward to examine the behaviour of these other proposals to the four-dimensional charged black holes  \reef{RNBas} or to the two-dimensional near-AdS$_2$ geometries \reef{AdScoord}. For the CV proposal, the techniques developed in \cite{Growth} are easily generalized to these new metrics and for the CV2.0 proposal, one need simply adapt the appropriate results for the bulk action. In either case, 
the late-time growth is found to be in keeping with the general expectations of eq.~\reef{general}, \ie $d{\cal C}/dt \sim S\,T$. 
Of course, these approaches to describing holographic complexity are only sensitive to the spacetime geometry and they would not be sensitive to the type of thermodynamic ensemble in question. However, it would be an interesting question to examine either of these proposals had a  well-motivated extension where the ensemble played a role in determining the behaviour of the holographic complexity. For instance, one might explore the role of boundary conditions in the context of the recent understanding of the holographic dual of the bulk symplectic form  \cite{Belin:2018fxe,Belin:2018bpg}.

The JT model provides a simple setup to study traversable wormholes \cite{dive,dive2} (see also \cite{Gao:2016bin}) and hence another interesting extension of the present work is to consider holographic complexity growth for traversable wormholes. In order to retrieve a quantum state which has fallen deep into the bulk, one would need to perform some operation which would prevent or reverse the natural tendency of the system to complexify. Perhaps it would be interesting to investigate the relation between the amount of quantum information that can be transmitted through the wormhole and the corresponding holographic complexity. Examining these ideas from the perspective of the Hayden-Preskill recovery protocol \cite{Hayden07,Yoshida:2017non} may also prove fruitful.

\section*{Acknowledgments}
We would like to thank Shira Chapman, Juan Hernandez, Robie Hennigar, Shan-Ming Ruan, Tokiro Numasawa, William Donnelly, Zachary Fisher, Ronak Soni and Alex Streicher for useful comments and discussions. We would also like to thank Adam Brown and Lenny Susskind for sharing with us a draft of their paper \cite{Brown:2018bms}. Research at Perimeter Institute is supported by the Government of Canada through Industry Canada and by the Province of Ontario through the Ministry of Research \& Innovation. RCM and BY are supported in part by Discovery Grants from the Natural Sciences and Engineering Research Council of Canada. RCM also received funding from the Simons Foundation through the ``It from Qubit'' collaboration. LQ would like to thank the Perimeter Scholars International and Perimeter Visiting Graduate Fellows programs for support during this research. The work of KG was supported in part by the JSPS Research Fellowship for Young Scientists. KG thanks Perimeter Institute for their hospitality during various stages of this project. RCM would also like to thank the KITP at UC Santa Barbara for their hospitality during the final stages of this project. At the KITP, this research was supported in part by the National Science Foundation under Grant No.~NSF PHY-1748958. 

\appendix

\section{More on Shock Waves}\label{App:Shock}

In this appendix, we discuss in more detail the calculation of the switchback effect in Reissner-Nordstrom-AdS background, as we presented in section \ref{subsec:Shocks}. The shock wave geometry is represented in the Penrose-like diagram in the right of figure \ref{PenroseHigherCharged}. This calculation below follows the analysis in \cite{Vad2}, so we refer the reader there for more details (as well as \cite{Vad1}).

The complexity dependence on $t_w$ can be determined by studying the time evolution of the special positions on the boundary of the WDW patch labeled by $r_b$, $r_s$, $r_m^1$ and $r_m^2$. From eq.~\eqref{eq:ShoPos}, we find that the time derivatives with respect to $t_L,\, t_R$ and $t_w$ are given by
\begin{align}
&\underline{r_s}: \, \qquad  \, \frac{d r_s}{d t_w} =- \frac{f_2 (r_s)}{2} \, , \qquad  \, \frac{d r_s}{d t_R} =- \frac{f_2 (r_s)}{2} \, , \qquad  \, \frac{d r_s}{d t_L} = 0 \, , \label{eq:DerCoord} \\
& \, \nonumber \\
&\underline{r_m^2}: \, \qquad  \, \frac{d r_m^2}{d t_w} = \frac{f_1 (r_m^2)}{2} \left[ 1 - \frac{f_2(r_s)}{f_1 (r_s)} \right] \, , \qquad  \, \frac{d r_m^2}{d t_R} = - \frac{f_1 (r_m^2)}{2} \frac{f_2 (r_s)}{f_1 (r_s)}   \, , \qquad  \, \frac{d r_m^2}{d t_L} = -\frac{f_1 (r_m^2)}{2} \, , \nonumber \\
& \, \nonumber \\
&\underline{r_b}: \, \qquad  \, \frac{d r_b}{d t_w} = - \frac{f_1 (r_b)}{2} \, , \qquad  \, \frac{d r_b}{d t_R} =0  \, , \qquad  \, \frac{d r_b}{d t_L} = \frac{f_1 (r_b)}{2} \, , \nonumber \\
& \, \nonumber \\
&\underline{r_m^1}: \, \qquad  \, \frac{d r_m^1}{d t_w} = \frac{f_2 (r_m^1)}{2} \left[ 1 - \frac{f_1(r_b)}{f_2 (r_b)} \right] \, , \qquad  \, \frac{d r_m^1}{d t_R} =\frac{f_2 (r_m^1)}{2}  \, , \qquad  \, \frac{d r_m^1}{d t_L} = \frac{f_2 (r_m^1)}{2} \frac{f_1 (r_b)}{f_2 (r_b)} \, . \nonumber
\end{align}

We will show next how to obtain the complexity of formation as a function of how early the shock wave is inserted (\ie of $t_w$). We will assume a dyonic black hole in four bulk dimensions and with a spherical horizon, as in the main text. Further, we only consider neutral shock waves, \ie the black hole charges before and after the shock wave remain equal. For convenience, we rewrite the metric here as
\begin{align}
&d s^2 = - F(r, v) d v^2 + 2 d r d v + r^2 (d\theta^2+\sin^2\theta\,d\phi^2)  \nonumber \\
&\text{with} \qquad \qquad F(r, v)= \frac{r^2}{L^2} + 1 - \frac{f_1(v)}{r} + \frac{q_e^2+q_m^2}{r^{2}} \, , \\
&\text{\ and} \qquad f_1 (v) = \omega_1 (1 - \mathcal{H}(v-v_s)) + \omega_2  \mathcal{H}(v-v_s) \, . \label{GeoApp} \\
\end{align}
The shock wave is then inserted at $v_s = - t_w$. 

\subsection*{Bulk Contribution}

The integrand of the bulk action (after integrating over the angular directions) can be written as 
\begin{equation}
\mathcal{I}_{b} (r) = \frac{4 \pi}{16 \pi G_N} \left( R - 2 \Lambda \right) - \frac{4 \pi}{g^2} \, F^2 =  \frac{1}{4 G_N} \left( - \frac{6}{L^2} + \frac{2  (q_{e}^2 - q_{m}^2 )}{r^{4}} \right)  \, .
\end{equation}
The bulk action in the Wheeler-DeWitt patch in figure \ref{PenroseHigherCharged} can be written as
\begin{align}
I^s_{bulk} =&  \int_{r_s}^{r_{\maxx}} d r \, r^{2} \, \mathcal{I}_{b} (r) \left( - 2 r^*_2(r) \right) + \int_{r_b}^{r_s}  d r \, r^{2} \, \mathcal{I}_{b} (r) \left( t_R + t_w \right)  \nonumber \\ 
& + \int_{r_{+, 1}}^{r_s} d r \, r^{2} \,  \mathcal{I}_{b} (r) (2 r_1^* (r_s) -2 r_1^* (r)  ) + 
\int_{r_m^2}^{r_{+,1}} d r \, r^{2} \, \mathcal{I}_{b} (r) \left( -t_L + t_w - 2 r_1^* (r) +  2 r_1^* (r_s)\right)
\nonumber \\
&  + \int_{r_{+,1}}^{r_{\maxx}} d r \, r^{2} \, \mathcal{I}_{b} (r)  \left( - 2 r^{*}_1 (r) \right)   +  \int_{r_b}^{r_{+, 1}} d r \, r^{2} \, \mathcal{I}_{b} (r)  \left( -t_w + t_L - 2 r^{*}_1 (r) \right)          \nonumber \\
& +  \int_{r_m^1}^{r_{b}} d r \, r^{2} \, \mathcal{I}_{b} (r)  \left( t_R + t_w - 2 r^{*}_2 (r) + 2 r_2^{*} (r_b) \right) \, .
\end{align}
We are interested in investigating the switchback effect, so as in section \ref{subsec:Shocks}, we set $t_L=t_R=0$ and probe the dependence on $t_w$, which is given by
\begin{align}
\frac{d I^s_{bulk} }{d t_w} &= - \frac{1}{2 G_N} \left[ \frac{r^3}{L^2} + \frac{q_{e}^2 - q_{m}^2}{r} \right] \bigg{|}_{r_m^1}^{r_s} - \frac{1}{2 G_N} \left[ \frac{r^3}{L^2} + \frac{q_{e}^2 - q_{m}^2}{r} \right] \bigg{|}_{r_m^2}^{r_b} \nonumber \\
&+\frac{1}{2 G_N} \left[ \frac{r^3}{L^2} + \frac{q_{e}^2 - q_{m}^2}{r} \right]  \frac{f_2 (r_s)}{f_1 (r_s)} \bigg{|}_{r_m^2}^{r_s} + \frac{1}{2 G_N} \left[ \frac{r^3}{L^2} + \frac{q_{e}^2 - q_{m}^2}{r} \right] \frac{f_1 (r_b)}{f_2 (r_b)}  \bigg{|}_{r_m^1}^{r_b} \, .
\end{align}
For large $t_w$, the derivative becomes
\begin{equation}
\frac{d I^s_{bulk} }{d t_w} \bigg{|}_{t_w \rightarrow \infty} = - \frac{1}{2 G_N} \left[ \frac{r^3}{L^2} + \frac{q_{e}^2 - q_{m}^2}{r} \right] \bigg{|}_{r_{-,1}}^{r_{+,1}} - \frac{1}{2 G_N} \left[ \frac{r^3}{L^2} + \frac{q_{e}^2 - q_{m}^2}{r} \right] \bigg{|}_{r_{-,2}}^{r_{+,2}}\, . \label{SBulktw}
\end{equation}

\subsection*{Joint and Counterterm Contributions}

We now evaluate the joint and counterterm contributions to the shock wave spacetime. In order to impose affine parametrization across the shock wave, we have the condition on the null normalization $\xi$ as \cite{Vad1, Vad2}
\begin{equation}
 \xi^{'} = \xi \, \frac{f_1 (r_s)}{f_2 (r_s)} \, , \qquad \, \qquad \xi^{''} = \xi \, \frac{f_2 (r_b)}{f_1 (r_b)} \, ,
\end{equation}
where $\xi$ is fixed at the boundaries with the usual prescription $k \cdot \partial_t |_{r \rightarrow \infty} = \pm \xi $, and the conditions for $\xi^{'}$ and $\xi^{''}$ are to ensure $\kappa=0$ for the null geodesic across the shock wave.
Because of the affine parametrization condition, the joints at $r_b$ and $r_s$ do not contribute \cite{Vad2}. The joints at $r_m^1$ and $r_m^2$ read
\begin{equation}
I_{\text{joint}} = \text{``UV terms"}  - \frac{1}{2 G_N} \left[ (r_m^1)^{2} \log \, \frac{|f_2 (r_m^1)|}{\xi \, \xi^{''}} + (r_m^2)^{2} \log \, \frac{|f_1 (r_m^2)|}{\xi \, \xi^{'}}   \right] \, .
\end{equation}

The counterterms associated to each of the null boundaries of the Wheeler-DeWitt patch (see \cite{Vad2} for more details), where $(\RN{1})$ refers to the the past null boundary that extends to the right asymptotic AdS boundary, $(\RN{2})$ to the future one that touches the left AdS boundary, $(\RN{3})$ to the past one that touches the left AdS boundary and finally $(\RN{4})$ to the future one that touches the right AdS boundary.
\begin{align}
 I_{\text{ct}}^{(\RN{1})} &= \text{``UV terms"}  - \frac{1}{2 G_N} (r_m^2)^{2} \left[ \log \left( \frac{2 \xi \, \lct }{r_m^2} \right) + \frac{1}{2} \right]  \nonumber \, \\
 &+ \frac{1}{2 G_N} \left( r_s^{2} -(r_m^2)^{2} \right) \log \frac{f_1 (r_s)}{f_2 (r_s)}  \nonumber  \, , \\
 I_{\text{ct}}^{(\RN{2})} &= \text{``UV terms"} - \frac{1}{2 G_N} (r_m^1)^{2} \left[ \log \left( \frac{2 \xi \, \lct }{r_m^1}\right) + \frac{1}{2} \right]   \\ 
 &+ \frac{1}{2 G_N} \left( r_b^{2} -(r_m^1)^{2} \right) \log \frac{f_2 (r_b)}{f_1 (r_b)}  \nonumber  \, , \\
 I_{\text{ct}}^{(\RN{3})} &= \text{``UV terms"}  - \frac{1}{2 G_N} (r_m^2)^{2} \left[ \log \left( \frac{2 \xi \, \lct }{r_m^2} \right) + \frac{1}{2} \right]  \nonumber \, \\
I_{\text{ct}}^{(\RN{4})} &= \text{``UV terms"}  - \frac{1}{2 G_N} (r_m^1)^{2} \left[ \log \left( \frac{2 \xi \, \lct }{r_m^1}\right) + \frac{1}{2} \right]  \nonumber \, 
\end{align}

If we add the joint and counterterm contributions, we find that the overall answer is independent of $\xi$,
\begin{align}
I_{ \text{joint} }  +  \sum I_{ \text{ct} } &= \text{``UV terms"} + \frac{1}{2 G_N} r_s^{2} \log \frac{f_1 (r_s)}{f_2 (r_s)} +  \frac{1}{2 G_N} r_b^{2} \log \frac{f_2 (r_b)}{f_1 (r_b)} \nonumber \\
& - \frac{1}{2 G_N} \, (r_m^1)^{2} \, \left[ \log \left( \frac{|f_2 (r_m^1)| 4\lct^2}{(r_m^1)^2} \right) + 1 \right] \nonumber \\
& - \frac{1}{2 G_N} \, (r_m^2)^{2} \, \left[ \log \left( \frac{|f_1 (r_m^2)| 4 \lct^2}{(r_m^2)^2} \right) + 1 \right] \, .
\end{align}
Then, for large $t_w$, the derivative simply reads
\begin{equation}
 \frac{d ( I_{ \text{joint} }  + \sum  I_{ \text{ct} } )}{d t_w} \bigg{|}_{t_w \rightarrow \infty} =  \frac{1}{2 G_N} \left[ \frac{r^3}{L^2} - \frac{q_e^2 + q_m^2}{r} \right] \bigg{|}^{r_{+,1}}_{r_{-,1}} + \frac{1}{2 G_N} \left[ \frac{r^3}{L^2} - \frac{q_e^2 + q_m^2}{r} \right]  \bigg{|}^{r_{+,2}}_{r_{-,2}} \, . \label{SJntstw}
\end{equation}

Combining equations eqs.~\eqref{SBulktw} and \eqref{SJntstw}, we obtain the simple result for the time derivative of the holographic with respect to $t_w$, at very early insertion times,
\begin{equation}
\frac{d \mathcal{C}_A}{d t_w} = \frac{1}{\pi G_N} \frac{q_e^2}{r} \bigg{|}_{r_{+,1}}^{r_{-,1}} +  \frac{1}{\pi G_N} \frac{q_e^2}{r} \bigg{|}_{r_{+,2}}^{r_{-,2}} \, . 
\end{equation}
Hence this derivative is directly proportional to $q_e^2$ for large $t_w$, such that if the black was purely magnetic, the derivative would vanish.

\subsubsection*{More on the switchback effect}

Let us evaluate more carefully the rate of complexity with respect to $t_w$ for times smaller than the scrambling time, such that $t_w < t^{*}_{\text{scr}} $. There is one big difference with respect to the switchback effect for Schwarzschild black holes as discussed in \cite{Vad2}, because now both the past and future boundaries of the Wheeler-DeWitt patch end at joints, instead of at a spacelike singularity. Therefore, there is always a transient term for small $T \, t_w$, that depends on $\ctL$, in analogy to the time evolution of the eternal black hole \cite{Growth}. In addition, the more magnetic charge that is present (\ie the smaller $\chi$ becomes), these transient effects will become important, which is the reason the curve in figure \ref{DyonLyapu} for small $\chi$  terminates long before the scrambling time, as these effects compete with each other.

The dominant contribution, for $\chi \gtrsim 1$, has a simple expression 
\begin{equation}
\frac{d \mathcal{C}_A}{d t_w} \simeq\ \mathcal{O} \left(\frac{\chi^2}{1+\chi^2} \, 
\frac{\qt^2}{\pi G_N}  \left(\frac{2}{r_{m}}  - \frac{1}{r_{s}} - \frac{1}{r_{b}}  \right) 
\epsilon e^{2 \pi T_1 t_w}
\right) +\ \text{transient}  \, .
\end{equation}
We denote $r_m$ the solution for meeting point equation in eq.~\eqref{MeetPntRN} with $t=0$, such that $r_m^1=r_m^2=r_m$. If we include the surface term in eq.~\eqref{mba2}, a similar expression holds 
\begin{equation}
\frac{d \mathcal{C}_A}{d t_w} \simeq\ \mathcal{O} \left(\frac{(1-\ap)\chi^2+\ap}{1+\chi^2} \, 
\frac{\qt^2}{\pi G_N}  \left(\frac{2}{r_{m}} - \frac{1}{r_s} - \frac{1}{r_b}  \right) \,  
\epsilon e^{2 \pi T_1 t_w} \right) +\  \text{transient}  \, .
\end{equation}

\section{Bulk Contribution from Maxwell Boundary Term} \label{MBAapp}

In section \ref{sec:MBA}, we argued that by using Stokes' theorem and the equations of motion, the Maxwell boundary term \eqref{mba2} could be written as the bulk contribution \eqref{MaxBdryActOnS}, and this identity was later used to simplify some of our calculations of the holographic complexity. However, there is clearly a subtlety: If one were to consider the case of a purely magnetic charge,  it is not hard to see that evaluating the boundary term \reef{mba2} on the boundaries of the WDW patch yields zero while the bulk integral on the right-hand side of eq.~\eqref{MaxBdryActOnS} is nonvanishing -- see details below. We elucidate the resolution of this inconsistency in the following. 

Consider the gauge potential for the magnetic charge by setting $q_e=0$ in  eq.~\eqref{FieldDyon} 
\beq
A= \frac{g q_m}{\sqrt{4 \pi G_N}} \,(1- \cos \theta) \, d \phi \equiv A_\mt{N} \, .  \label{FieldMagnetic}
\eeq
Of course, we recognize that this choice is not well-defined at $\theta =\pi$. As a result, this problem is inherited by the integrand appearing in the boundary term, \ie 
\begin{equation}\label{naive}
F^{\mu\nu}A_\nu=\frac{g^2}{4\pi G_N}\,\frac{q_m^2}{r^4 \sin{\theta}}\, (1-\cos{\theta})\ \delta^\mu_\theta\,,
\end{equation}
is singular at $\theta=\pi$. Hence we can not properly apply Stokes' theorem in the way that was implicitly done in deriving eq.~\eqref{MaxBdryActOnS}.

We can evade this problem by describing the gauge potential on two patches, one covering the north pole ($\theta=0$) and the other covering the south pole ($\theta=\pi$). For example, on a(n open) hemisphere $H^2_\mt{N}$ covering $0\leq\theta<\pi/2$, we use the potential $A_\mt{N}$ in eq.~\reef{FieldMagnetic}, while on the complementary hemisphere $H^2_\mt{S}$ which covers the south pole with $\pi/2<\theta \leq \pi$, we define
\beq
A_\mt{S} = -\frac{g}{\sqrt{4 \pi G_N}} \, q_m (1+\cos \theta) \, d \phi \, . \label{ASplit}
\eeq
Since the two gauge potentials and the corresponding integrands (as in eq.~\reef{naive}) are well-defined on their respective patches, we can apply Stoke's theorem in each patch separately but now the resulting boundary terms now include an integral over the surface $\theta=\pi/2$ where the contributions involving $A_\mt{N}$ and $A_\mt{S}$ do {\it not} cancel. 

Let us explicitly illustrate this point for the Reissner-Nordstrom-AdS black holes discussed in section \ref{sec:RNBHS} but in the case, where the charge is purely magnetic. Given the form of the metric \reef{RNBas}, the WDW patch has the form of a direct product
${\cal M}={\cal N}\times S^2$. Now setting $q_e=0$ in eq.~\reef{FieldDyon}, the field strength becomes $F=\frac{g\,q_m}{\sqrt{4\pi G_N}}\,\sin\theta d\phi \wedge d\theta$ and so evaluating the right-hand side of eq.~\reef{MaxBdryActOnS} yields \begin{equation}\label{poult3}
  \frac{1}{2g^2}\int_{\mathcal{M}} d^{4} x \sqrt{-g}\, F^{\mu \nu} F_{ \mu \nu}
 = \frac{\ap\,q_m^2}{G_N}\ \int_{\cal N} \frac{dt\, dr}{r^2} \, .
\end{equation}
Next we turn to the boundary term as given in eq.~\reef{mba2}. Certainly if we only integrate over the boundary of the WDW patch $\partial {\cal M}=\partial {\cal N}\times S^2$, then the result is zero since as we saw in eq.~\reef{naive}, the combination $F^{\mu\nu}A_\nu$ only has a $\theta$ component. However, as we discussed above, we should actually integrate over the boundaries of all of the patches where the gauge potential is well-defined, \eg using the prescription outlined above, the boundary integral runs over
\beq 
(\partial {\cal M})'=(\partial {\cal N}\times S^2) \ \cup\ ({\cal N}\times \partial H^2_\mt{N})\ \cup\ ({\cal N} \times \partial H^2_\mt{S})\,.
\label{poult4}
\eeq
Hence if we integrate over this boundary, eq.~\reef{mba2} yields
\beqa
\frac{\ap}{g^2} \int_{(\partial \mathcal{M})'} \!\!\! d \Sigma_{\mu} \, F^{\mu \nu} \, A_{\nu}
&=&\frac{\ap}{g^2} \int_{{\cal N}\times \partial H^2_\mt{N}} \!\!\!\!\!
d \Sigma_{\mu} \, F^{\mu \nu} \, (A_\mt{N}-A_\mt{S})_{\nu}
\nonumber\\
&=& \frac{\ap\,q_m^2}{G_N}\ \int_{\cal N} \frac{dt\, dr}{r^2} \, .
\label{poult5}
\eeqa
where in the first line, we have used $\partial H^2_\mt{S}=-\, \partial H^2_\mt{N}$ where the sign indicates the two boundaries have opposite orientations. Hence with the prescription that the Maxwell boundary term
\reef{mba2} is integrated on the boundaries of all of the patches used to define the gauge potential, we see that the bulk and boundary integrals precisely match, \ie eqs.~\reef{poult3} and \reef{poult5} yield exactly the same result.

Hence the lesson that we take away here is that when we introduce the boundary term \reef{mba2} to evaluate the WDW action for a magnetically charged system, we should understand that this term is not only integrated over the (geometric) boundary of the WDW patch but rather it is integrated over the boundaries of all of the patches introduced to produce a properly defined gauge potential everywhere.\footnote{We might add that the various surfaces between the two patches also play the role of a boundary in deriving the equations of motion. In particular, the gauge potentials in two neighbouring patches cannot be varied independently, but rather one imposes a boundary condition fixing their difference, \eg in the example above, $\delta(A_\mt{N}-A_\mt{S})=0$ on $\partial H^2_\mt{S}=-\, \partial H^2_\mt{N}$ (when $\gamma=0$).}

\end{document}